\PassOptionsToPackage{table,xcdraw}{xcolor}
\PassOptionsToPackage{hyphens}{url}
\documentclass[sigconf]{acmart}
\renewcommand\footnotetextcopyrightpermission[1]{}

\setcopyright{none} % No copyright notice required for submissions

\usepackage{comment}
\usepackage{color}
\usepackage{algorithm}
\usepackage[noend]{algpseudocode}
\usepackage{amsmath}
\usepackage{balance}
\usepackage{caption}
\usepackage{subfig}
\usepackage{graphicx}
\usepackage{acronym}
\usepackage{url}
\usepackage{hyperref}
\usepackage{caption}
\usepackage{pifont}
\usepackage[table,xcdraw]{xcolor}
\usepackage[normalem]{ulem}
\useunder{\uline}{\ul}{}
\usepackage{siunitx}

\newcounter{TODO}

\newcommand{\nip}[1]{\vspace{1ex}\noindent\textbf{#1}}

\newcommand{\sysname}{\textsc{Truda}}

\acrodef{FL}{Federated Learning}
\acrodef{FFL}{Framework for Federated Learning}
\acrodef{AI}{artificial intelligence}
\acrodef{ML}{machine learning}
\acrodef{MLaaS}{machine learning as a service}
\acrodef{DL}{deep learning}
\acrodef{DNN}{deep neural network}
\acrodef{ConvNet}{convolutional neural network}
\acrodef{SGX}{Software Guard Extensions}
\acrodef{SEV}{Secure Encrypted Virtualization}
\acrodef{EVM}{encrypted virtual machine}
\acrodef{TDX}{Trust Domain Extensions}
\acrodef{TEE}{Trusted Execution Environment}
\acrodef{MEE}{Memory Encryption Engine}
\acrodef{OS}{operating system}
\acrodef{SoC}{System-on-Chip}
\acrodef{SP}{Secure Processor}
\acrodef{PEF}{Protected Execution Facility}
\acrodef{HPVS}{Hyper Protect Virtual Servers}
\acrodef{FedSGD}{Federated Stochastic Gradient Descent}
\acrodef{FedAvg}{Federated Averaging}
\acrodef{IoT}{Internet of Thing}
\acrodef{VM}{virtual machine}
\acrodef{SME}{Secure Memory Encryption}
\acrodef{AES}{Advanced Encryption Standard}
\acrodef{ASID}{Address Space Identifier}
\acrodef{MSE}{Mean Squared Error}
\acrodef{OVMF}{Open Virtual Machine Firmware}
\acrodef{IoT}{Internet of Things}

\acrodef{PDH}{Platform Diffie-Hellman Public Key}
\acrodef{GODH}{Guest Owner Diffie-Hellman Public Key}
\acrodef{CEK}{Chip Endorsement Key}
\acrodef{ASK}{AMD SEV Signing Key}
\acrodef{ARK}{AMD Root Key} 
\acrodef{OCA}{Owner Certificate Authority} 
\acrodef{TIK}{Transport Integrity Key}
\acrodef{TEK}{Transport Encryption Key}
\acrodef{KEK}{Key Encryption Key}
\acrodef{KIK}{Key Integrity Key}
\acrodef{VEK}{VM Encryption Key}

\acrodef{DHKE}{Diffie-Hellman Key Exchange}
\acrodef{HE}{Homomorphic Encryption}
\acrodef{SMC}{Secure Multi-Party Computation}
\acrodef{DP}{Differential Privacy}
\acrodef{CDP}{Centralized Differential Privacy}
\acrodef{LDP}{Local Differential Privacy}
\acrodef{DDP}{Distributed Differential Privacy}

\acrodef{DLG}{Deep Leakage from Gradients}
\acrodef{iDLG}{Improved Deep Leakage from Gradients}
\acrodef{IG}{Inverting Gradients}

\algrenewcommand\algorithmicrequire{\textbf{Precondition:}}
\algrenewcommand\algorithmicensure{\textbf{Postcondition:}}
\def\HiLi{\leavevmode\rlap{\hbox to \hsize{\color{gray!50}\leaders\hrule height .8\baselineskip depth .5ex\hfill}}}
  
\usepackage{afterpage}
\usepackage{multicol}
\newsavebox{\shortpagebox}
\makeatletter
\newcommand{\shortpage}[1]% #1= \twocolumn text to wrap into \onecolumn page
{\par
  \setbox\shortpagebox=\vbox{\strut #1\par}%
  \afterpage{\onecolumn
    \begin{multicols}{2}
    \unvbox\AP@partial
    \end{multicols}}%
  \unvbox\shortpagebox
\par}
\makeatother

\begin{document}
\title{Separation of Powers in Federated Learning}
\author{Pau-Chen Cheng}
\affiliation{%
\institution{IBM Research}
\state{New York}
\country{USA}
}
\email{pau@us.ibm.com}
\author{Kevin Eykholt}
\affiliation{%
\institution{IBM Research}
\state{New York}
\country{USA}
}
\email{kheykholt@ibm.com}
\author{Zhongshu Gu}
\authornote{Corresponding authors.}
\affiliation{%
\institution{IBM Research}
\state{New York}
\country{USA}
}
\email{zgu@us.ibm.com}
\author{Hani Jamjoom}
\affiliation{%
\institution{IBM Research}
\state{New York}
\country{USA}
}
\email{jamjoom@us.ibm.com}
\author{K. R. Jayaram}
\authornotemark[1]
\affiliation{%
\institution{IBM Research}
\state{New York}
\country{USA}
}
\email{jayaramkr@us.ibm.com}
\author{Enriquillo Valdez}
\affiliation{%
\institution{IBM Research}
\state{New York}
\country{USA}
}
\email{rvaldez@us.ibm.com}
\author{Ashish Verma}
\affiliation{%
\institution{IBM Research}
\state{New York}
\country{USA}
}
\email{ashish.verma1@ibm.com}

\begin{abstract}
\ac{FL} enables collaborative training among mutually distrusting parties. Model updates, rather than training data, are concentrated and fused in a central aggregation server. 
A key security challenge in \ac{FL} is that an untrustworthy or compromised aggregation process might lead to unforeseeable information leakage. This challenge is especially acute due to recently demonstrated attacks that have reconstructed large fractions of training data from ostensibly ``sanitized'' model updates.

In this paper, we introduce \sysname{}, a new cross-silo \ac{FL} system, employing a trustworthy and decentralized aggregation architecture to break down information concentration with regard to a single aggregator.
Based on the unique computational properties of model-fusion algorithms, all exchanged model updates in \sysname{} are disassembled at the parameter-granularity and re-stitched to random partitions designated for multiple TEE-protected aggregators. Thus, each aggregator only has a fragmentary and shuffled view of model updates and is oblivious to the model architecture. Our new security mechanisms can fundamentally mitigate training data reconstruction attacks, while still preserving the final accuracy of trained models and keeping performance overheads low.
\end{abstract}
\maketitle
\pagestyle{plain}
\section{Introduction}
\label{sec:introduction}
\acf{FL}~\cite{mcmahan2017communication} provides a collaborative training mechanism, which allows multiple parties to build a joint \ac{ML} model.
\ac{FL} allows parties to retain private data within their controlled domains. Only model updates are shared to a central aggregation server.
The security setting of \ac{FL} is especially attractive for mutually distrusting/competing training participants as well as holders of sensitive data (e.g., health and financial data) seeking to preserve data privacy. 

\ac{FL} is \emph{typically} deployed in two scenarios: \emph{cross-device} and \emph{cross-silo}~\cite{kairouz2019advances}. 
The \emph{cross-device} scenario involves a large number of parties ($>1000$), but each party has a small number of data items, constrained compute capability, and limited energy reserve (e.g., mobile phones or IoT devices). They are highly unreliable and are expected to drop and join frequently. Examples include a large organization learning from data stored on employees' devices and a device manufacturer training a model from private data located on millions of its devices (e.g., Google Gboard~\cite{bonawitz2019towards}). A \emph{trusted authority}, which performs aggregation and orchestrates training, is typically present in a \emph{cross-device} scenario. Contrarily, in the \emph{cross-silo} scenario, the number of parties is small, but each party has extensive compute capabilities (with stable access to electric power and/or equipped with hardware \ac{ML} accelerators) and large amounts of data. The parties have reliable participation throughout the entire \ac{FL} training life-cycle, but are more susceptible to sensitive data leakage. Examples include multiple hospitals collaborating to train a tumor detection model on radiographs, multiple banks collaborating to train a credit card fraud detection model, etc. In \emph{cross-silo} scenarios, there exists \emph{no presumed central trusted authority}. All parties involved in the training are \emph{equal} collaborators. The deployments often involve hosting aggregation in public clouds, or alternatively one of the parties acting as, and providing infrastructure for aggregation. In this paper, we focus on examining the trustworthiness of collaborative learning in \emph{cross-silo} scenarios.

There was a \emph{misconception} that the exchanged model updates in \ac{FL} communications would contain 
far less, if any, information about the raw training data. Thus, sharing model updates was considered to 
be ``privacy-preserving.'' However, although not easily discernible, training data information is still embedded in the model updates. Recent research~\cite{melis2019exploiting, zhu2019deep,zhao2020idlg, geiping2020inverting,yin2021see} 
has demonstrated the feasibility and ease of inferring private attributes and reconstructing large fractions
of training data by exploiting model updates, thereby challenging the privacy promises of \ac{FL}
in the presence of honest-but-curious aggregation servers.
Furthermore, since aggregation often runs on untrustworthy cloud or third-party infrastructures in the \emph{cross-silo} scenarios, we need to re-examine the trust model and system architecture of current \ac{FL} frameworks under this new attack scenario.

Existing solutions that reinforce \ac{FL} privacy include (1) \ac{DP} based aggregation through the addition of statistical noise to model updates~\cite{mcmahan2017learning, geyer2017differentially, shokri2015privacy, bhowmick2018protection, truex2019hybrid}, and  (2) using \ac{SMC} or \ac{HE}~\cite{aono2017privacy,hardy2017private,bonawitz2017practical, mohassel2017secureml} for aggregation. Both techniques have several drawbacks. The former often significantly decreases the accuracy of the trained model and needs careful hyper-parameter tuning to minimize accuracy loss. The latter is computationally expensive, with aggregation overheads significantly outpacing training time~\cite{jayaram-cloud2020}. 

Our work is motivated by the following key insights: (I) The \emph{concentration} of model updates in a central \ac{FL} aggregator discloses significantly more information than what is required by the aggregation/model-fusion algorithms. This gap, indeed, facilitates data reconstruction attacks if the central aggregator is compromised. (II) \ac{FL} model-fusion algorithms are \emph{bijective} and only involve \emph{coordinate-wise} arithmetic operations across model updates. Partitioning and (internally) shuffling model updates do not change the fusion results, thus having no impact on the final model accuracy and convergence rate compared to traditional (insecure) \ac{FL} training. We only need to ensure that the local transformation of model updates is deterministic, reversible, and synchronized across parties. From the attacker's point of view, partitioning and shuffling operations fully disrupt the completeness and data-order of model updates, which are indispensable for reconstructing training data. (III) \acp{TEE} can help establish confidential and remote-attestable execution entities on untrustworthy servers. The missing link is how to authenticate and connect all individual entities to bootstrap trust in a distributed \ac{FL} ecosystem. 

In this paper, we introduce \sysname{}\footnote{\sysname{} stands for \underline{TRU}stworthy and \underline{D}ecentralized \underline{A}ggregation}, a new cross-silo \ac{FL} system for \ac{DNN} training. In \sysname{}, we employ multiple, rather than one, \ac{TEE}-protected decentralized aggregators. All model updates are disassembled at the parameter-granularity and re-stitched (with shuffling) to random partitions designated for different aggregators. Thus, each aggregator only has a fragmentary and out-of-order view of each model update and is oblivious to the model architecture. This new system architecture can fundamentally minimize the information leakage surface for data reconstruction attacks, but with no utility loss regarding model aggregation. We have implemented three composable security-reinforced mechanisms in \sysname{}:

\nip{Trustworthy Aggregation.} 
We leverage \acp{TEE} to protect the model-fusion process. Every aggregator runs within an \ac{EVM} via AMD \ac{SEV}. All in-memory data are kept encrypted at runtime during model aggregation. To bootstrap trust between parties and aggregators, we design a two-phase attestation protocol and develop a series of tools for integrating/automating confidential computing in \ac{FL}. Each party can authenticate trustworthy aggregators before participating in \ac{FL} training. End-to-end secure channels, from the parties to the \acp{EVM}, are established after attestation to protect model updates in transit.

\nip{Decentralized Aggregation.} 
Decentralization was primarily investigated in distributed learning~\cite{bonawitz2019towards, vanhaesebrouck2017decentralized, bellet2018personalized, koloskova2019decentralized, lalitha2019peer} as a load balancing technique for enhancing system performance;
model updates were distributed among aggregator replicas with each update assigned to a single replica and each replica seeing the entire model update from a party. This still provided a fertile ground for data reconstruction attacks on the aggregator side.
Instead, we choose a security-centric decentralization strategy by employing fine-grained model partitioning to break down information concentration. We launch multiple aggregators within \acp{TEE}. Each aggregator only receives a fraction of model
update with no knowledge of the whole model architecture. The parties share a model-mapper at parameter-granularity for disassembling and re-stitching a local model update into disjoint partitions, which are dispatched to corresponding aggregators. Furthermore, users can deploy multiple aggregators to physical servers at different geo-locations and potentially with diversified \acp{TEE}, e.g., Intel \acs{SGX}~\cite{mckeen2013innovative}/ \acs{TDX}~\cite{tdx2020intel}, IBM \acs{PEF}~\cite{hunt2021confidential}, etc. Thus, we can prevent model aggregation from becoming a single point of failure (i.e., leaking entire and intact model updates) under security attacks. 

\nip{Shuffled Aggregation.} Based on the bijective property of model-fusion algorithms, we allow parties to permute the fragmentary model updates to further obfuscate the information sent to each aggregator. The permutation changes dynamically at each training round. This strategy guarantees that even if \emph{all} decentralized aggregators are breached, adversaries will still require the permutation key to be able to decipher the correct ordering of the model updates and reconstruct the training data.

In our security analysis, we reproduced three state-of-the-art training data reconstruction attacks~\cite{zhu2019deep, zhao2020idlg, geiping2020inverting} and plugged in \sysname{} for generating the model updates. Our experiments demonstrate that \sysname{} renders all the attacks ineffective for reconstructing local training data. In our performance evaluation, we measured the accuracy/loss and latency for training deep learning models on the datasets, \emph{MNIST}, \emph{CIFAR-10}, and \emph{RVL-CDIP}\cite{harley2015icdar}, with a spectrum of aggregation algorithms and \ac{FL} configurations. We demonstrate that \sysname{} can achieve the same level of accuracy/loss and converge at the same rate, with minimal performance overheads compared to the traditional federated learning platform as baseline. 
\section{Threat Model}
\label{sec:threatmodel}
Our threat model assumes honest-but-curious aggregation servers, which are susceptible to compromise. 
Adversaries attempt to attack aggregators and inspect model updates uploaded from parties. Their purpose is to reconstruct training data of parties that participate in the \ac{FL} training.
We consider that the parties involved in \ac{FL} training are benign and will not collude with other parties to share training data.
This threat model is the same as in the \ac{FL} data reconstruction attacks~\cite{zhu2019deep, zhao2020idlg, geiping2020inverting,yin2021see}. 

In addition, our threat model is consistent with the one assumed in AMD \ac{SEV}. We consider that parties involved in \ac{FL} trust AMD \ac{SoC} and the \acp{EVM} launched to hold the model aggregation workloads. Adversaries can not only execute user-level code on the aggregator's hosting machines, but can also execute malicious code at the level of privileged system software, e.g., \ac{OS}, hypervisor, BIOS, etc. The attacker may also have physical access to the DRAM of hosting machines. Our current \ac{FL} system implementation is based on the 1st-generation \ac{SEV}~\cite{kaplan2016amd} on EPYC 7642 (ROME) microprocessor, which is the latest release available on the market. Based on the white papers, the following \ac{SEV} generations, i.e., SEV-ES~\cite{kaplan2017protecting} and SEV-SNP~\cite{sev2020strengthening},  will include confidentiality protection for \ac{VM} register state and integrity protection to defend against memory corruption, aliasing, remapping, and replay attacks. Our threat model can be elevated to follow the stronger isolation protection of future \ac{SEV} releases. 

Recently, some research efforts~\cite{werner2019severest,buhren2019insecure,li2019exploiting,li2020crossline,wilke2020sevurity} focus on discovering potential vulnerabilities of \ac{SEV}. In this paper, we do not intend to address these problems and consider they will be fixed with the AMD's firmware updates or in the upcoming SEV-SNP release. However, \emph{even if \acp{EVM} are breached}, our design of decentralized and shuffled aggregation of model updates will still be effective at preventing adversaries from reconstructing training data.
\section{Background}
\label{sec:background}

This section provides relevant information on common \ac{FL} aggregation algorithms, data reconstruction attacks, computational properties of \ac{FL} aggregation algorithms and AMD \ac{SEV} that are leveraged by \sysname{}.

\subsection{Aggregation Algorithms for DNN Training}
\ac{FedSGD}~\cite{shokri2015privacy} and \ac{FedAvg}~\cite{mcmahan2017communication} are the most common \ac{FL} aggregation algorithms for \ac{DNN} training, employing iterative merging and synchronizing model updates. Other \ac{DNN} aggregation methods, e.g., Byzantine-robust fusions like \emph{Coordinate Median}\cite{yin2018byzantine}/\emph{Krum}\cite{blanchard2017machine} and \emph{Paillier crypto fusion}\cite{aono2017privacy,liu2019secure,truex2019hybrid}, have the similar algorithmic structure with additional security enhancements. \sysname{} can support all of them with no change. Here we describe the algorithms of \emph{\ac{FedSGD}} and \emph{\ac{FedAvg}} in detail.

We use $\theta$ to denote model parameters and $L$ for the loss function. Each party has its own training data/label pairs $(x_i, y_i)$. The parties choose to share the gradients $\nabla_\theta L_\theta (x_i, y_i)$ for a data batch to the aggregator. The aggregator computes the gradient sum of all parties and lets the parties synchronize their model parameters: $
\theta \leftarrow \theta - \eta \sum_{i=1}^{N} \nabla_\theta L_\theta (x_i, y_i)$.
This aggregation algorithm is called \emph{\ac{FedSGD}}. Alternatively, the parties can also train for several epochs locally and upload the model parameters: $\theta^{i} \leftarrow \theta^{i} - \eta \nabla_{\theta^{i}} L_{\theta^{i}} (x_i, y_i)$
to the aggregator. The aggregator computes the weighted average of model parameters $\theta \leftarrow \sum_{i=1}^{N} \frac{n_i}{n} \theta^{i}$, where $n_i$ is the size of training data on party $i$ and $n$ is the sum of all $n_i$. Then, the aggregator sends the aggregated model parameters back to the parties for synchronization. This aggregation algorithm is called \emph{\ac{FedAvg}}. 
\emph{\ac{FedAvg}} and \emph{\ac{FedSGD}} are equivalent if we train only one batch of data in a single \ac{FL} training round and synchronize model parameters, as gradients can be computed from the difference of two successive model parameter uploads.  
As \emph{\ac{FedAvg}} allows parties to batch multiple SGD iterations before synchronizing updates, it would be more challenging for data reconstruction attacks to succeed.

\subsection{Data Reconstruction Attacks} 
Exchanging model updates in \emph{\ac{FedAvg}} and \emph{\ac{FedSGD}} was considered privacy-preserving as original training data were not directly included in communications. However, recent attacks, e.g., \ac{DLG}~\cite{zhu2019deep}, \ac{iDLG}~\cite{zhao2020idlg}, and \ac{IG}~\cite{geiping2020inverting}, have demonstrated that it is possible to derive training data samples from the model updates. 

In \ac{DLG}~\cite{zhu2019deep}, the attack reconstructed a training sample $x$ based on the shared gradient updates. The attack randomly initialized a dummy input $x'$ and label $y'$, which were fed into the model in order to compute the loss gradients. Then, the attack used an L-BGFS solver to minimize the following cost function in order to reconstruct $x$:
\begin{align*}
    \underset{x', y'}{\operatorname{argmin}} || \nabla_{\theta}L_{\theta}(x',y') - \nabla_{\theta}L_{\theta}(x,y) ||^{2}
\end{align*}
As the differentiation requires second order derivatives, the attack only works on models that are twice differentiable. 
Zhao et al.~\cite{zhao2020idlg} found that although the \ac{DLG} attack was effective, the reconstructions and labels generated after optimization were sometimes of low quality and incorrect respectively. In their \ac{iDLG} attack, the authors demonstrated that the signs of the loss gradients with respect to the correct label are always opposite to the signs of the other labels. Thus, the ground truth labels can be inferred based on the model updates, which improve reconstruction quality. 

\ac{IG}~\cite{geiping2020inverting} makes two major modifications to \ac{DLG} and \ac{iDLG}. First, the authors asserted that it is not the magnitudes of the gradients that are important, but rather the directions of the gradients. Based on this reasoning, they used a cosine distance cost function, which encouraged the attack to find reconstructions that resulted in the same changes in gradients' directions. Their new cost function is:
\begin{align*}
    \underset{x' \in [0,1]}{\operatorname{argmin}} \;1- \frac{\langle \nabla_{\theta}L_{\theta}(x',y), \nabla_{\theta}L_{\theta}(x,y)\rangle}{||\nabla_{\theta}L_{\theta}(x',y)|||| \nabla_{\theta}L_{\theta}(x,y)||} + \alpha TV(x')
\end{align*}
They also (i) constrained their search space to $[0,1]$, (ii) added total variation as an image prior, and (iii) minimized their cost function based on the signs of the loss gradients and the ADAM optimizer. This modification was inspired by adversarial attacks on \acp{DNN}, which used a similar technique to generate adversarial inputs~\cite{szegedy2013intriguing}. 

\subsection{Bijectivity of Aggregation Algorithms}
Most \acp{DNN} aggregation algorithms (e.g., \emph{\ac{FedAvg}}, \emph{\ac{FedSGD}}, \emph{Krum}, \emph{Coordinate Median}, \emph{Paillier crypto fusion}, etc.) 
only involve bijective summation and averaging operations. In simple terms, if a model is represented as a flattened vector $M$, these algorithms perform coordinate-wise fusion across parties. That is, they add or average the elements at $M$[$i$] from all parties --- parameters at a given index $i$ can be fused with no dependency on those at any other indices. 

By examining the methods of data reconstruction attacks, we observe that a \emph{global} view of model updates is required in the attacks' optimization procedures. The completeness and data-order of model updates are crucial for reconstructing training data. Lack of either will lead to reconstruction failures. On the contrary, as \ac{FL} aggregation algorithms for \ac{DNN} training only involve coordinate-wise operations at the parameter granularity, data completeness and ordering are not required. 

Therefore, we are able to partition an entire model update into multiple pieces, deploy them to multiple servers, and execute the same fusion algorithms independently across all servers. Furthermore, before aggregation, each partitioned vector can also be shuffled at each training round, as long as all parties permute in the same order. Parties can reverse the permutation and merge the aggregated partitions locally when they receive the aggregated model updates for synchronization.

\subsection{AMD SEV}
\ac{SEV}~\cite{kaplan2016amd} is a confidential computing technology introduced by AMD. It aims to protect security-sensitive workloads in public cloud environments. \ac{SEV} depends on AMD \ac{SME} to enable runtime memory encryption. Along with the AMD Virtualization (AMD-V) architecture, \ac{SEV} can enforce cryptographic isolation between guest \acp{VM} and the hypervisor. Therefore, \ac{SEV} can prevent privileged system administrators, e.g., at the hypervisor level, from accessing the data within the domain of an \ac{EVM}.

When \ac{SEV} is enabled, \ac{SEV} hardware tags all code and data of a \ac{VM} with an \ac{ASID}, which is associated with a distinct ephemeral \ac{AES} key, called \ac{VEK}. The keys are managed by the AMD \ac{SP}, which is a 32-bit ARM Cortex-A5 micro-controller integrated within the AMD \ac{SoC}. Runtime memory encryption is performed via on-die memory controllers. Each memory controller has an \ac{AES} engine that encrypts/decrypts data when it is written to main memory or is read into the \ac{SoC}. The control over memory page encryption is via page tables. Physical address bit 47, a.k.a., \emph{C-bit}, is used to mark whether the memory page is encrypted. 

Similar to other \acp{TEE}, \ac{SEV} also provides a remote attestation mechanism for authenticating hardware platforms and attesting \acp{EVM}. The authenticity of the platform is proven with an identity key signed by AMD and the platform owner. Before provisioning any secrets, \ac{EVM} owners should verify both the authenticity of \ac{SEV}-enabled hardware and the measurement of \ac{OVMF}, which enables UEFI support for booting \acp{EVM}.
\section{System Design}
\label{sec:design}
\begin{figure*}[!ht]
\centering
\includegraphics[width=1\textwidth]{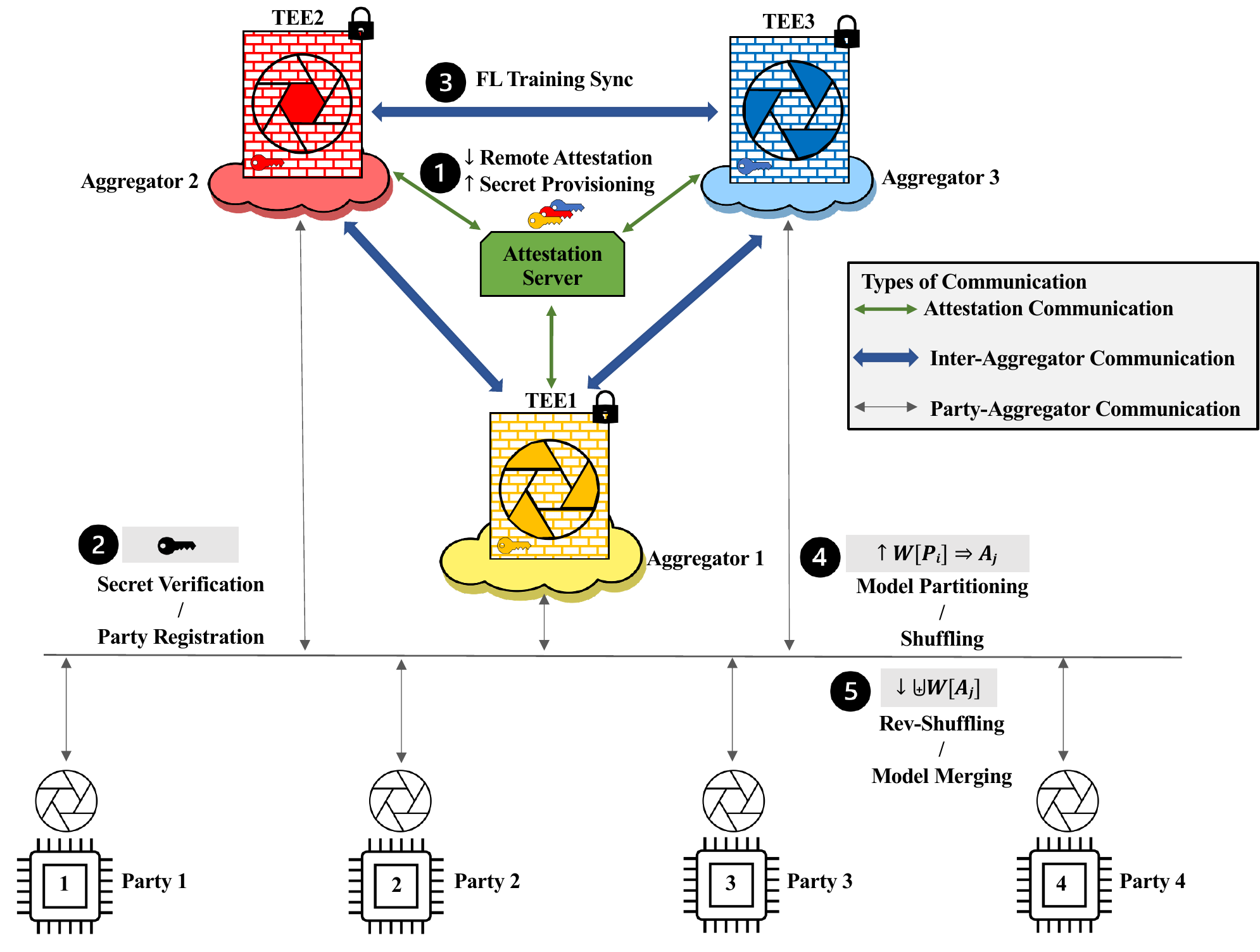}
\caption{System Architecture of \sysname{}}
\label{fig:arch}
\end{figure*}
In this section, we detail the design of \sysname{} and demonstrate how it effectively mitigates information leakage channels for \ac{FL} data reconstruction attacks. We describe three key security mechanisms of \sysname{}: (1) enabling trustworthy \ac{FL} aggregation with a two-phase attestation protocol, (2) separating a central aggregator to multiple decentralized entities, each with only a fragmentary view of the model updates, and (3) shuffling model updates dynamically for each training round. These three mechanisms are composable and can operate with no mutual dependency. Such a multi-faceted design makes \sysname{} more resilient when facing unanticipated security breaches. For example, assuming that memory encryption of \acp{TEE} is broken due to some new zero-day vulnerabilities, the adversaries will still not be able to reconstruct training data from the model updates (leaked from \acp{TEE}) as they are fragmentary and obfuscated with mechanisms (2) and/or (3) enabled. 

We give a concrete deployment example of \sysname{} in Figure~\ref{fig:arch} and discuss the workflow step by step. Similar to traditional \ac{FL}, in \sysname{}, each party needs to register with the aggregators to participate in the training. One aggregator initiates the training process by notifying all parties. During training, aggregators engage in a number of training rounds with all parties. At each training round, each party uses its local training data to produce a new model update and upload it to the aggregators. The aggregators merge model updates from all parties and dispatch the aggregated version back to all parties. The global training ends once pre-determined training criteria (like number of epochs or accuracy) are met.  

Different from traditional \ac{FL}, \sysname{}'s deployment involves multiple aggregators running within \acp{TEE}, rather than a single central aggregator. Aggregators need to communicate with each other for training synchronization. In addition, we also deploy an attestation server as a trusted anchor, which is responsible for attesting the aggregator's workload and provisioning secrets on behalf of all parties.  
\begin{figure*}[!ht]
\centering
\includegraphics[width=1\textwidth]{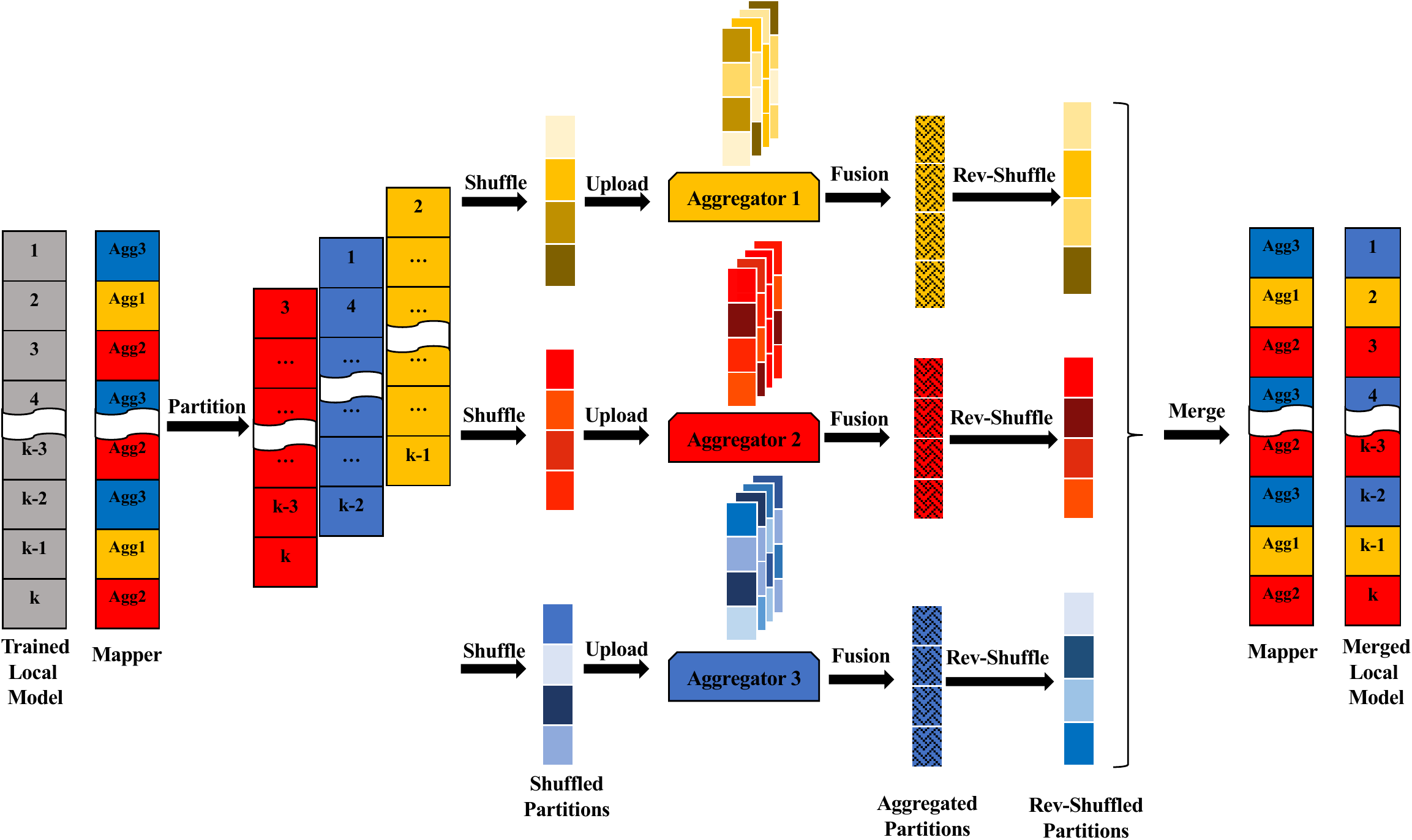}
\caption{Model Partitioning and Dynamic Permutation}
\label{fig:partition}
\end{figure*}
\subsection{Trustworthy Aggregation}
As mentioned earlier, model updates exchanged between parties and aggregators may contain essential information for reverse engineering private training data. We need to eliminate the channels for adversaries to intercept and inspect model updates in transit and also in use. In our design, we enforce cryptographic isolation for \ac{FL} aggregation via \ac{SEV}. The aggregators execute within \acp{EVM}. Each \ac{EVM}'s memory is protected with a distinct ephemeral \ac{VEK}. Therefore, we can protect the confidentiality of model aggregation from unauthorized users, e.g., system administrators, and privileged software running on the hosting servers. 
AMD provides attestation primitives for verifying the authenticity of individual \ac{SEV} hardware/firmware. We design a new attestation protocol upon the primitives to bootstrap trust between parties and aggregators in the distributed \ac{FL} setting. This \ac{FL} attestation protocol consists of two phases: 

\nip{Phase 1: Launching Trustworthy Aggregators.} 
First, we need to securely launch \ac{SEV} \acp{EVM} with aggregators running within. To establish the trust of \acp{EVM}, attestation must prove that (1) the platform is an authentic AMD \ac{SEV}-enabled hardware providing the required security properties, and  
(2) the \ac{OVMF} image to launch the \ac{EVM} is not tampered. Once the remote attestation is completed, we can provision a secret, as a unique identifier of a trustworthy aggregator, to the \ac{EVM}. The secret is injected into \ac{EVM}'s encrypted physical memory and used for aggregator authentication in Phase 2.   

In Figure~\ref{fig:arch}, Step \ding{182} shows an attestation server that facilitates remote attestation. The \ac{EVM} owner instructs the AMD \ac{SP} to export the certificate chain from the \ac{PDH} down to the \ac{ARK}. This certificate chain can be verified by the AMD root certificates. The digest of \ac{OVMF} image is also included in the attestation report along with the certificate chain.   

The attestation report is sent to the attestation server, which is provisioned with the AMD root certificates. The attestation server verifies the certificate chain to authenticate the hardware platform and check the integrity of \ac{OVMF} firmware. Thereafter, the attestation server generates a launch blob and a \ac{GODH} certificate. They are sent back to the \ac{SP} on the aggregator's machine for negotiating a \ac{TEK} and a \ac{TIK} through \ac{DHKE} and launching the \acp{EVM}.  

We can retrieve the \ac{OVMF} runtime measurement through the \ac{SP} by pausing the \ac{EVM} at launch time. We send this measurement (along with the \ac{SEV} API version and the \ac{EVM} deployment policy) to the attestation server to prove the integrity of UEFI booting process. Only after verifying the measurement, the attestation server generates a packaged secret, which includes an ECDSA prime251v1 key. The hypervisor injects this secret into the \ac{EVM}'s physical memory space as a unique identifier of a trusted aggregator and continue the  launching process. Our secret injection procedure follows the 1st-generation \ac{SEV}'s remote attestation protocol. With the upcoming \ac{SEV}-SNP, AMD \ac{SP} can also measure and attest the layout of \ac{EVM}'s initial memory and its contents~\cite{sev2020strengthening}. Thus, the new \ac{SEV}-SNP's attestation mechanism can further reinforce the integrity of the launching process. 

\nip{Phase 2: Aggregator Authentication.}
Parties participating in \ac{FL} must ensure that they are interacting with trustworthy aggregators with runtime memory encryption protection. To enable aggregator authentication, in Phase 1, the attestation server provisions an ECDSA key as a secret during \ac{EVM} deployment. This key is used for signing challenge requests and thus serves to identify a legitimate aggregator. In step \ding{183} of Figure~\ref{fig:arch}, before participating in FL, a party first attests an aggregator by engaging in a challenge response protocol. The party sends a randomly generated nonce to the aggregator. The aggregator digitally signs the nonce using its corresponding ECDSA key and then returns the signed nonce to the requesting party. The party verifies that the nonce is signed with the corresponding ECDSA key. If the verification is successful, the party then proceeds to register with the aggregator to participate in FL. This process is repeated for all aggregators.

After registration, end-to-end secure channels can be established to protect communications between aggregators and parties for exchanging model updates. We enable TLS to support mutual authentication between a party and an aggregator. Thus, all model updates are protected both within \acp{EVM} and in transit.

\subsection{Decentralized Aggregation}
Enabling trustworthy aggregation alone may not be sufficient. We cannot expect that \acp{TEE} are omnipotent and that there will be no security vulnerabilities discovered in the future. In fact, we have already observed some security breaches with regard to \ac{SEV}~\cite{werner2019severest,buhren2019insecure,li2019exploiting,li2020crossline,wilke2020sevurity}. Therefore, we enhance our design to ensure that even if \acp{TEE} are breached, adversaries cannot reconstruct training data from model updates.

Decentralization has been primarily explored (e.g., Bonawitz et al.~\cite{bonawitz2019towards}) in distributed learning as a load balancing technique to address the performance problems with respect to a central server, and to scale to a large number of parties. This has typically involved distributing parties among aggregator replicas (``party-partitioning''), and
the replicas co-ordinating among themselves to aggregate intermediate results. In this setting, each model update reaches
an aggregator replica in its entirety, making reverse engineering possible. In other research on fully-decentralized/peer-to-peer distributed learning\cite{vanhaesebrouck2017decentralized, bellet2018personalized, koloskova2019decentralized, lalitha2019peer}, there exists no central aggregator. Each party exchanges model updates with their neighbors and merges the received updates with its local model. But this scheme cannot prevent information leakage either as the entire model updates are still circulated among parties in the network and susceptible to reconstruction attacks. 

In \sysname{}, we choose a different decentralization strategy with model partitioning enabled. Each aggregator only receives a fraction of each model update and runs within an \ac{EVM}. For example, in Figure~\ref{fig:arch}, we establish three aggregators. Each party authenticates and registers with all aggregators respectively. 

\nip{Inter-Aggregator Training Synchronization.} We maintain communication channels between aggregators for training synchronization, e.g., step \ding{184} of Figure~\ref{fig:arch}. Any one of the aggregators can start the training and become the \emph{initiator} node by default. All the other aggregators become \emph{follower} nodes and wait for the commands from the \emph{initiator}. At each training round, the \emph{initiator} first notifies all parties to start local training and retrieve the model updates for fusion. Thereafter, it notifies all the \emph{follower} nodes to pull their corresponding model updates, aggregate them together, and distribute the aggregated updates back to the parties.      

\nip{Randomized Model Partitioning.}
Due to the bijective nature of \ac{FL} fusion algorithms, we can split a model update into disjoint partitions based on the number of available aggregators. Before training starts, we randomly generate a model-mapper for each \ac{DNN} model (to be trained). We allow the parties to choose the proportion of model parameters for each aggregator. This model-mapper is shared by all the parties that participate in the \ac{FL} training. For example, in Figure~\ref{fig:partition}, the $k$ parameters of a local model are mapped to three aggregators. The colors imply the aggregator attributes for each parameter within the model. The model update is disassembled and rearranged at the parameter-granularity for different aggregators (step \ding{185} in Figure~\ref{fig:arch}). Once parties receive fused model updates from different aggregators, they query the model-mapper again to merge model updates to its original positions within the local model (step \ding{186} in Figure~\ref{fig:arch}). A shared model-mapper can be generated either by (i) using a consensus algorithm like Raft, or (ii) constructed at each party using random number generators seeded by the same random value (e.g., from League of Entropy\footnote{\url{https://blog.cloudflare.com/league-of-entropy/}}).

Decentralized aggregation significantly defends against the illegitimate leak of model information at the aggregation point as aggregators no longer store the complete model updates nor the model architectures. As we demonstrate in Section~\ref{sec:security}, even missing a very small fraction of a model update totally renders the data reconstruction attacks ineffective. Thus, decentralized aggregation requires adversaries to compromise all \ac{TEE}-protected aggregators (which can be isolated in different protected domains) and obtain the model-mapper (protected within party's training devices) to piece together complete model updates in their original order. 

\subsection{Shuffled Aggregation}
To further obfuscate the information transferred from the parties to the aggregators, we employ a dynamic permutation scheme to shuffle the partitioned model updates at every training round. Each permutation is seeded by the \emph{combination} of a permutation key (e.g., dispatched from a trusted key generation server) agreed among all parties and a dynamically generated training identifier synchronized at the start of each training round. Thus, the permutation changes every training round, but is deterministic across all parties. After parties receive aggregated model updates, they rev-shuffle the results back to their original order as shown in Figure~\ref{fig:partition}.

The dynamic permutation scheme is based on the insight that the data-order of model updates is irrelevant for model-fusion algorithms, while it is crucial for optimization procedures used in data reconstruction attacks. With dynamic permutation enabled, adversaries only obtain obfuscated model updates and the data order dynamically changes at each training round. If the permutation seed is not leaked, it is infeasible for adversaries to reconstruct any training data even if they compromise all \ac{TEE}-protected aggregators. In addition, this shuffled aggregation mechanism works in the deployments with either a central aggregator (as in traditional \ac{FL}) or multiple decentralized aggregators (as in \sysname{}). 
\begin{table*}[!h]
\caption{Comparison of Fidelity Threshold (MSE) for DLG and iDLG with Model Partitioning and Permutation}

\label{tab:dlgidlg}
\begin{tabular}{lccc|ccc|ccc|ccc}
\cline{2-13}
 &
  \multicolumn{3}{c|}{DLG (w/o perm)} &
  \multicolumn{3}{c|}{iDLG (w/o perm)} &
  \multicolumn{3}{c|}{DLG (with perm)} &
  \multicolumn{3}{c}{iDLG (with perm)} \\ \cline{2-13}  &
  \multicolumn{3}{c|}{Partition \%} &
  \multicolumn{3}{c|}{Partition \%} &
  \multicolumn{3}{c|}{Partition \%} &
  \multicolumn{3}{c}{Partition \%} \\ \hline

\multicolumn{1}{l}{\cellcolor[HTML]{C0C0C0}Fidelity Threshold (MSE)} &
  \multicolumn{1}{|c}{{\ul \textbf{100}}} &
  \multicolumn{1}{c}{60} &
  \multicolumn{1}{c|}{20} &
  \multicolumn{1}{c}{{\ul \textbf{100}}} &
  \multicolumn{1}{c}{60} &
  \multicolumn{1}{c|}{20} &
  \multicolumn{1}{c}{100} &
  \multicolumn{1}{c}{60} &
  \multicolumn{1}{c|}{20} &
  \multicolumn{1}{c}{100} &
  \multicolumn{1}{c}{60} &
  \multicolumn{1}{c}{20} \\ \hline
\rowcolor[HTML]{FD6864} 
\multicolumn{1}{l|}{\cellcolor[HTML]{C0C0C0}{[}0, \num{1.0e-03})} &
  {\ul \textbf{66.6\%}} &
  0.0\% &
  0.0\% &
  {\ul \textbf{83.7\%}} &
  0.0\% &
  0.0\% &
  0.0\% &
  0.0\% &
  0.0\% &
  0.0\% &
  0.0\% &
  0.0\% \\ \hline
\rowcolor[HTML]{9AFF99} 
\multicolumn{1}{l|}{\cellcolor[HTML]{C0C0C0}{[}\num{1.0e-03}, \num{1.0e-02})} &
  {\ul \textbf{0.8\%}} &
  0.0\% &
  0.0\% &
  {\ul \textbf{1.2\%}} &
  0.0\% &
  0.0\% &
  0.0\% &
  0.0\% &
  0.0\% &
  0.0\% &
  0.0\% &
  0.0\% \\ \hline
\rowcolor[HTML]{9AFF99} 
\multicolumn{1}{l|}{\cellcolor[HTML]{C0C0C0}{[}\num{1.0e-02}, \num{1.0e-01})} &
  {\ul \textbf{0.3\%}} &
  0.0\% &
  0.0\% &
  {\ul \textbf{0.1\%}} &
  0.0\% &
  0.0\% &
  0.0\% &
  0.0\% &
  0.0\% &
  0.0\% &
  0.0\% &
  0.0\% \\ \hline
\rowcolor[HTML]{9AFF99} 
\multicolumn{1}{l|}{\cellcolor[HTML]{C0C0C0}{[}\num{1.0e-01}, \num{1.0e+00})} &
  {\ul \textbf{0.2\%}} &
  0.0\% &
  0.0\% &
  {\ul \textbf{0.2\%}} &
  0.0\% &
  0.0\% &
  0.0\% &
  0.0\% &
  0.0\% &
  0.0\% &
  0.0\% &
  0.0\% \\ \hline
\rowcolor[HTML]{9AFF99} 
\multicolumn{1}{l|}{\cellcolor[HTML]{C0C0C0}{[}\num{1.0e+00}, \num{1.0e+02})} &
  {\ul \textbf{8.1\%}} &
  38.9\% &
  20.5\% &
  {\ul \textbf{6.6\%}} &
  66.5\% &
  18.3\% &
  0.0\% &
  0.0\% &
  0.2\% &
  0.2\% &
  0.2\% &
  0.7\% \\ \hline
\rowcolor[HTML]{9AFF99} 
\multicolumn{1}{l|}{\cellcolor[HTML]{C0C0C0}$\geq$ \num{1.0e+02}} &
  {\ul \textbf{24.0\%}} &
  61.1\% &
  79.5\% &
  {\ul \textbf{8.2\%}} &
  33.5\% &
  81.7\% &
  100.0\% &
  100.0\% &
  99.8\% &
  99.8\% &
  99.8\% &
  99.3\% \\ \hline
\end{tabular}
\end{table*}

\begin{table}[!h]
\caption{Comparison of Final Cosine Distance for IG with Model Partitioning and Permutation}
\label{tab:ig}
\begin{tabular}{lccc|ccc}
\cline{2-7}
 &
  \multicolumn{3}{c|}{IG (partition)} &
  \multicolumn{3}{c}{IG (partition+perm)} \\ \cline{2-7} 
%\rowcolor[HTML]{EFEFEF} 
\cellcolor[HTML]{FFFFFF}{\color[HTML]{333333} } &
  \multicolumn{3}{c|}{Partition \%} &
  \multicolumn{3}{c}{Partition \%} \\ \hline
%\rowcolor[HTML]{EFEFEF} 
\multicolumn{1}{l|}{\cellcolor[HTML]{C0C0C0}Cosine Distance}   & {\ul \textbf{100}}   & 60    & 20   & 100   & 60    & 20    \\ \hline
\rowcolor[HTML]{FD6864} 
\multicolumn{1}{l|}{\cellcolor[HTML]{C0C0C0}{[}0, 0.01)}   & {\ul \textbf{100\%}} & 0\%   & 0\%  & 0\%   & 0\%   & 0\%   \\ \hline
\rowcolor[HTML]{9AFF99} 
\multicolumn{1}{l|}{\cellcolor[HTML]{C0C0C0}{[}0.01, 0.2)} & {\ul \textbf{0\%}}   & 0\%   & 0\%  & 0\%   & 0\%   & 0\%   \\ \hline
\rowcolor[HTML]{9AFF99} 
\multicolumn{1}{l|}{\cellcolor[HTML]{C0C0C0}{[}0.2, 0.4)}  & {\ul \textbf{0\%}}   & 100\% & 0\%  & 0\%   & 0\%   & 0\%   \\ \hline
\rowcolor[HTML]{9AFF99} 
\multicolumn{1}{l|}{\cellcolor[HTML]{C0C0C0}{[}0.4, 0.6)}  & {\ul \textbf{0\%}}   & 0\%   & 98\% & 0\%   & 0\%   & 0\%   \\ \hline
\rowcolor[HTML]{9AFF99} 
\multicolumn{1}{l|}{\cellcolor[HTML]{C0C0C0}{[}0.6, 0.8)}  & {\ul \textbf{0\%}}   & 0\%   & 2\%  & 0\%   & 0\%   & 0\%   \\ \hline
\rowcolor[HTML]{9AFF99} 
\multicolumn{1}{l|}{\cellcolor[HTML]{C0C0C0}{[}0.8, 1{]}}  & {\ul \textbf{0\%}}   & 0\%   & 0\%  & 100\% & 100\% & 100\% \\ \hline
\end{tabular}
\end{table}
\section{Implementation}
\label{sec:implementation}
We developed \sysname{} by extending the publicly available IBM \ac{FFL}~\cite{ludwig2020ibm} to support trustworthy aggregation, decentralized multi-aggregators with model partitioning, and dynamic permutation of model updates. We containerized the aggregator application and employed Kata Container~\cite{kata} to deploy aggregator containers inside lightweight \acp{EVM}. 
We used an AMD EPYC 7642 (ROME) microprocessor running \ac{SEV} API Version 0.22~\cite{sevapi}. We extended QEMU with Feldman-Fitzthum's patch~\cite{qemupatch}\footnote{This patch will be included in the release of QEMU 6.0} to support AMD \ac{SEV} \emph{LAUNCH\_SECRET} and extended Kata-runtime to provide remote attestation via client-side gRPC~\cite{grpc} communication with the attestation server. Finally, We implemented our attestation server as a gRPC service using a modified version of the AMD \ac{SEV}-Tool~\cite{sevtool} to support \ac{EVM} owners' tasks, e.g., attesting the AMD \ac{SEV}-enabled hardware platform, verifying the \ac{OVMF} launch measurement, and generating the launch blob.

\section{Security Analysis}
\label{sec:security}
We evaluated the effectiveness of \sysname{} against three \ac{FL} attacks that attempted to reconstruct training data based on model updates: \ac{DLG}~\cite{zhu2019deep}, \ac{iDLG}~\cite{zhao2020idlg}, and  \ac{IG}~\cite{geiping2020inverting}. We used the implementations from their public git repositories~\cite{DLG-git,iDLG-git,IG-git} for our experiments. First, we evaluated each attack with only model partitioning enabled, varying the partition factor by $40\%$, thus lowering the percentage of model updates accessible to the attack. A partition factor of $100\%$ means the attack had access to the entire model update. Then, we enabled dynamic permutation together with model partitioning and re-evaluated the model, again varying the partition factor by $40\%$.

The current design of \sysname{} does not allow aggregators to maintain a global model, thus they have no knowledge of the model architecture. \ac{DLG}, \ac{iDLG}, and \ac{IG} can neither retrieve the unmodified model updates associated with an input sample nor query the model for the dummy input's current loss gradients. As such, in a real deployment of \sysname{}, these attacks would not succeed as they lack both these two critical components. However, to analyze the effects of the security measures of \sysname{}, we relaxed the constraints and allowed adversaries to query the complete, unperturbed model as a blackbox. Therefore, the attacks could compute the dummy inputs' loss gradients, but the original inputs' loss gradients were still transformed by \sysname{}. In this stronger attack scenario, we demonstrate that \sysname{} still remains effective and prevents the attacks from leaking information through reconstructing data from model updates.

\subsection{DLG and iDLG Results}

We used a randomly initialized \emph{LeNet} model for evaluation as done by prior works~\cite{zhu2019deep, zhao2020idlg} and evaluated both attacks using $1000$ randomly selected inputs from the \emph{CIFAR-100} dataset. \emph{CIFAR-100} is a dataset with $32\times32$ color images in $100$ classes. We ran each attack for 300 iterations. The effectiveness of \sysname{} against \ac{DLG} and \ac{iDLG} is reported in Table \ref{tab:dlgidlg}. 

\begin{figure*}[!ht]
\centering
\includegraphics[width=0.9\textwidth]{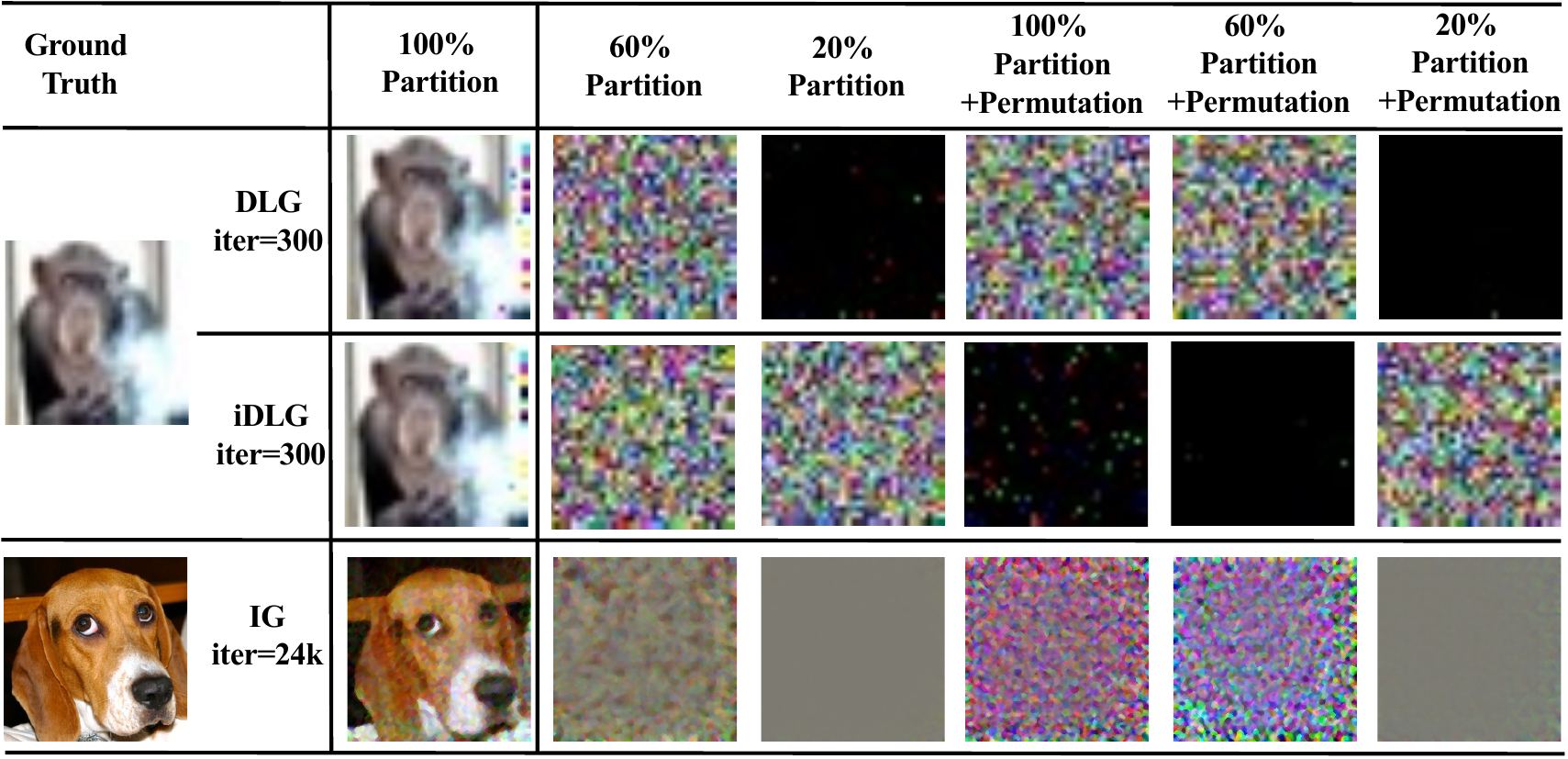}
\caption{Reconstruction Examples of DLG/iDLG/IG with Model Partitioning and Permutation}
\label{fig:reconstructexamples}
\end{figure*}

We partitioned the results into six ranges based on the \ac{MSE} of each image. \ac{MSE} is the metric adopted in \ac{DLG}/\ac{iDLG} for measuring the quality of reconstructed images in \emph{CIFAR-100}. Through visual inspection, an \ac{MSE} below \num{1.0e-03} compared to the original images resulted in recognizable reconstructions. We highlight this threshold in red in Table~\ref{tab:dlgidlg}. Without \sysname{} in place, \ac{DLG} and \ac{iDLG}, resulted in generating $66.6\%$ and $83.7\%$ recognizable reconstructions respectively with a partition factor of $100\%$ (i.e., no model partitioning). These results are used as the baseline and highlighted in the two \emph{underlined} columns. However, as soon as model partitioning is enabled, the reconstruction quality drops significantly. Due to model partitioning, both attacks' estimates of the original gradients are increasingly inaccurate as fewer of the original model's weights are available. In turn, both attacks cannot correctly minimize the cost function, which we observed during the attack process. With only a $60\%$ partition rate, neither attack is able to generate any recognizable reconstructions. Enabling dynamic permutation in addition to model partitioning adds an additional layer of protection against reconstruction attacks as evidenced by the increased \ac{MSE} of the reconstructions in Table \ref{tab:dlgidlg}. Even with all of the model weights, neither attack is able to generate an recognizable reconstruction as dynamic permutation of the model weights prevents the attacks from correctly aligning their gradient estimations. In the first and second rows of Figure \ref{fig:reconstructexamples}, we present the reconstructions generated by the \ac{DLG} and \ac{iDLG} attacks without and with \sysname{} enabled for one of the images. In Appendix~\ref{sec:exampleappendix}, we include another example (Figure~\ref{fig:dlgidlgexamples}) with more details of intermediate iterations. 

\subsection{IG Results}
We used a randomly initialized \emph{ResNet-18} model for evaluation as done by prior work~\cite{geiping2020inverting} and evaluated \ac{IG} using $50$ randomly selected inputs from the \emph{ImageNet} dataset. \emph{ImageNet} is a dataset with $224\times224$ color images in $1000$ classes. We ran the attack for $24,000$ iterations with two random restarts.

The authors of \ac{IG} remarked that the reconstruction quality and amount of information leaked are highly dependent on the images. \ac{MSE} is no longer an accurate metric for measuring image similarity for large-sized data samples of \emph{ImageNet}. Instead, we measured the \emph{cosine distance}, which is used as the cost function of \ac{IG}, to show that \sysname{} effectively hinders the optimization procedure. We partitioned the cosine distance (bounded in $[0,1]$) into six ranges and present the results in Table~\ref{tab:ig}. Without \sysname{} in place, the cosine distance of \ac{IG}'s cost function is always smaller than $0.01$ with a partition factor of $100\%$ (i.e., no model partitioning). These results are used as the baseline and highlighted in the column with the underlined values. With \sysname{} enabled, \ac{IG} can no longer correctly minimize the cost function. The cosine distance values in the optimization procedures stuck at the level significantly larger than $0.01$. For example, with a $60\%$ partition rate and no permutation, all cosine distance values are in the range of $[0.2, 0.4)$. With permutation enabled, the cosine distance is further increased to the range of $[0.8, 1]$. Visual inspection of the results reveals that \ac{IG} cannot generate any recognizable reconstructions with \sysname{}'s partitioning and permutation in place. In the third row of Figure~\ref{fig:reconstructexamples}, we present the reconstructions generated by the \ac{IG} attack without and with \sysname{} enabled for one of the images. In Appendix~\ref{sec:exampleappendix}, we include five more reconstruction examples for \ac{IG} in Figure~\ref{fig:igexamples}. 
\section{Performance Evaluation}
\label{sec:performance}
We evaluated the performance of \sysname{} with two metrics. First, we measured the loss/accuracy of the models generated at each training round. It demonstrates that the convergence rate of \sysname{} is aligned with the base system and \sysname{} does not lead to model accuracy degradation. Second, we measured the latency of \ac{FL} training. The latency of model training refers to the total time to finish a specified number of training rounds. We recorded the time after finishing each training round at the aggregator. The latency results reflect the performance overhead incurred by the security features added in \sysname{}. 

Our evaluation covers a spectrum of \emph{cross-silo} \ac{FL} training applications from three aspects: (1) adaptability to different \ac{FL} aggregation algorithms, (2) performance comparisons with different numbers of participating parties, and (3) support for larger \ac{DNN} models with non-IID training data distribution. 

In our evaluation environment, each party ran within a \ac{VM} in a datacenter. We assigned each \ac{VM} with 16 cores of Intel Xeon E5-2690 CPU, one Nvidia Tesla P100 GPU, 120 GB DRAM; the \ac{OS} is Redhat Enterprise Linux 7.0-64. We set up three aggregators to run within the \ac{SEV} \acp{EVM}. The aggregators ran on a machine with AMD EPYC 7642 CPU; the host OS is Ubuntu 20.04 LTS. The baseline for comparison is IBM \ac{FFL} with one central aggregator. The party's configurations, model architectures, and hyper-parameter settings are the same as for \sysname{} and \ac{FFL}. 

\subsection{Training with Different Fusion Algorithms}
\begin{figure*}[!ht]
\centering
\subfloat[Loss/Accuracy Comparison: Iterative Averaging]{
\includegraphics[width=0.33\textwidth]{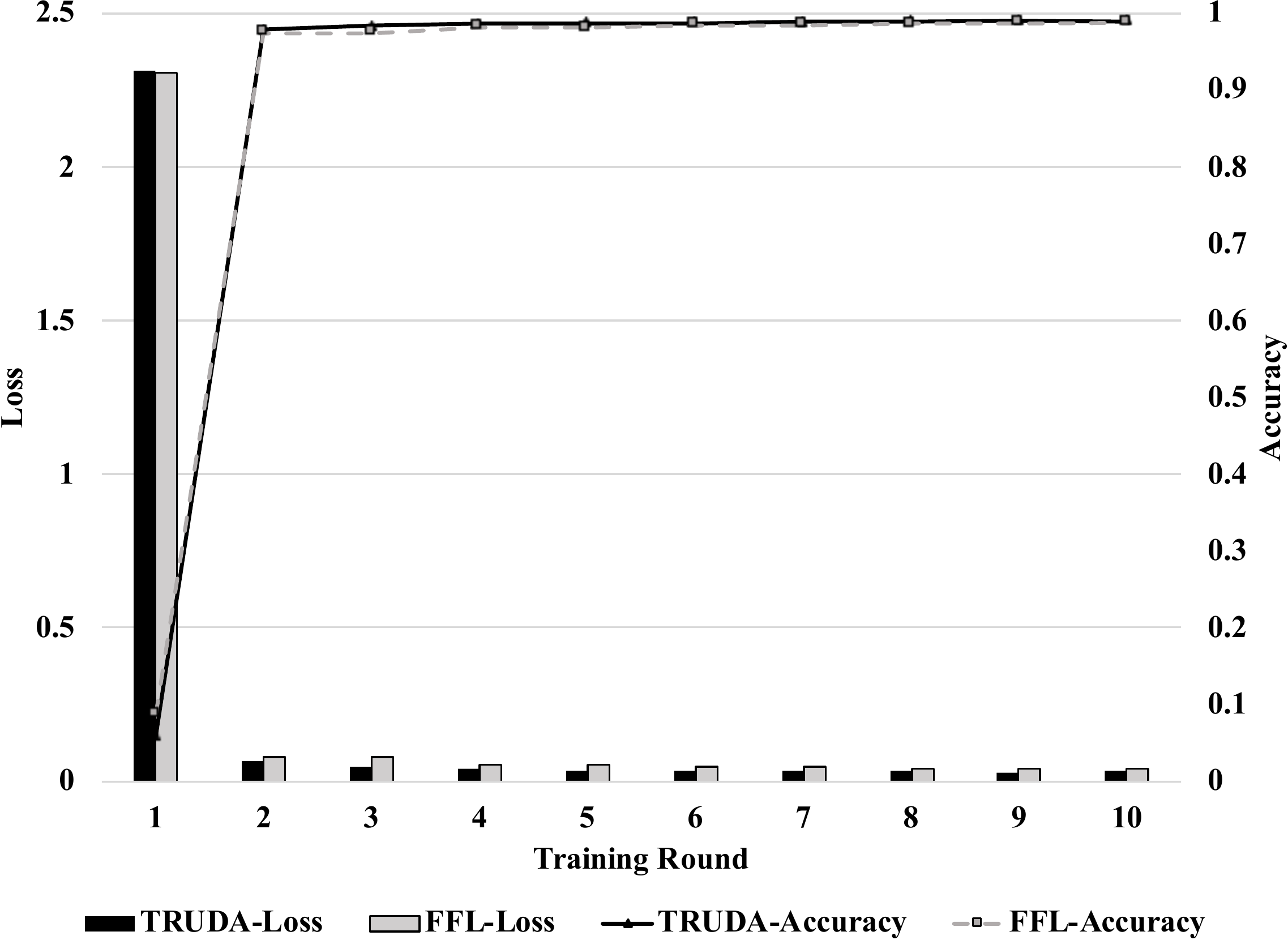}
\label{fig:mnistiterlossacc}}
\subfloat[Loss/Accuracy Comparison: Coordinate Median]{
\includegraphics[width=0.33\textwidth]{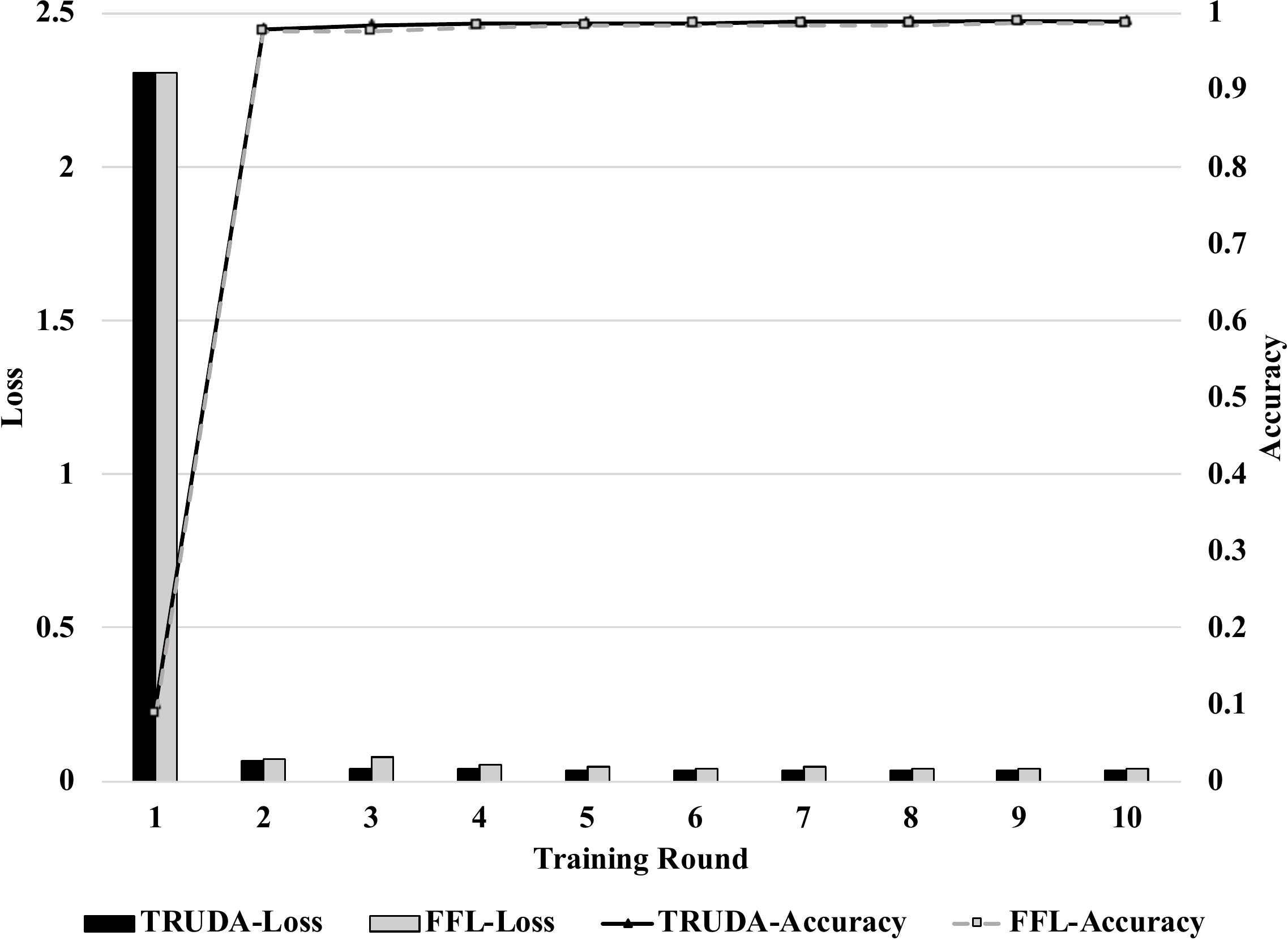}
\label{fig:mnistmedianlossacc}}
\subfloat[Loss/Accuracy Comparison: Paillier crypto system]{
\includegraphics[width=0.33\textwidth]{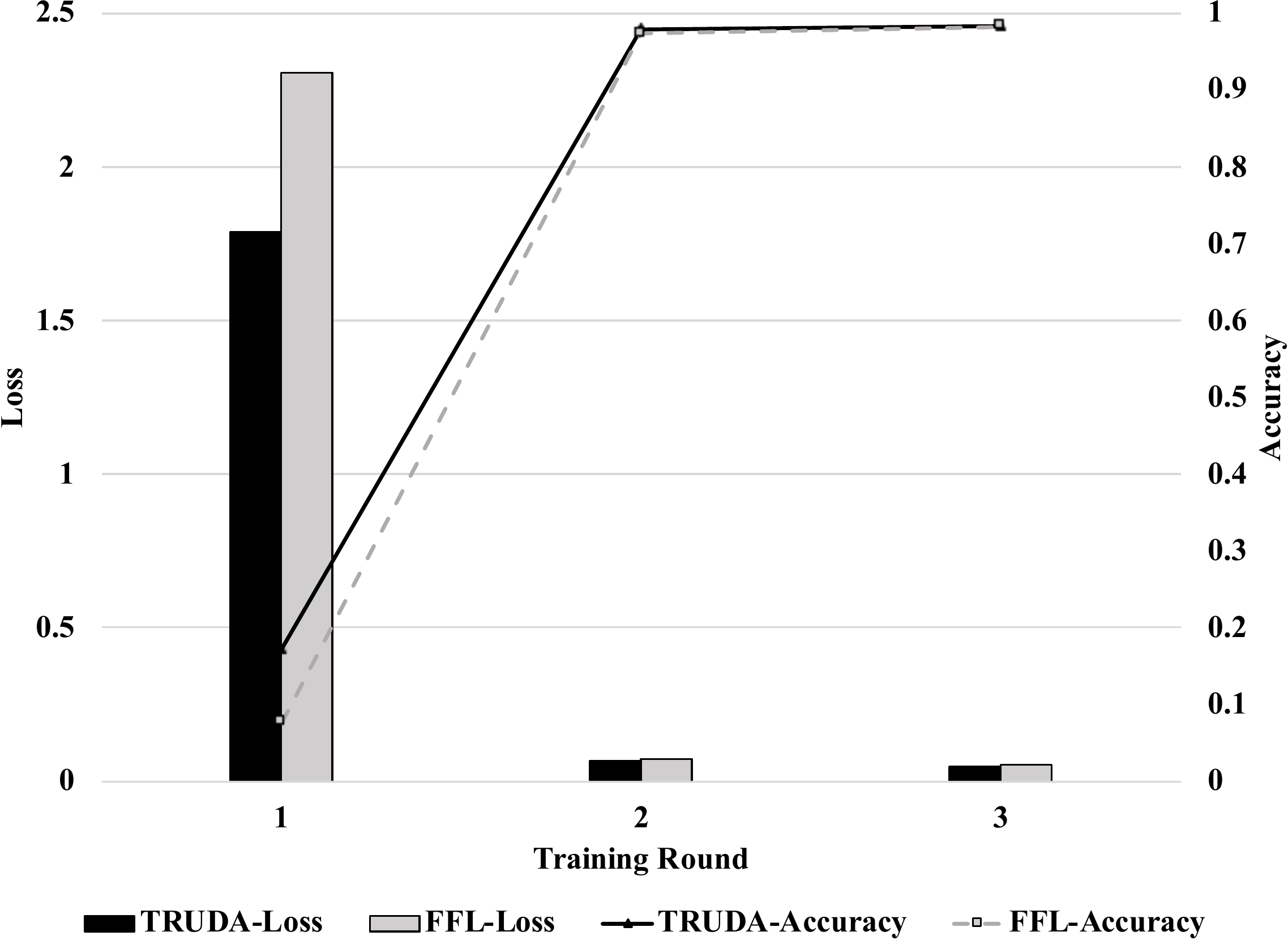}
\label{fig:mnistpaillierlossacc}}

\subfloat[Latency Comparison: Iterative Averaging]{
\includegraphics[width=0.33\textwidth]{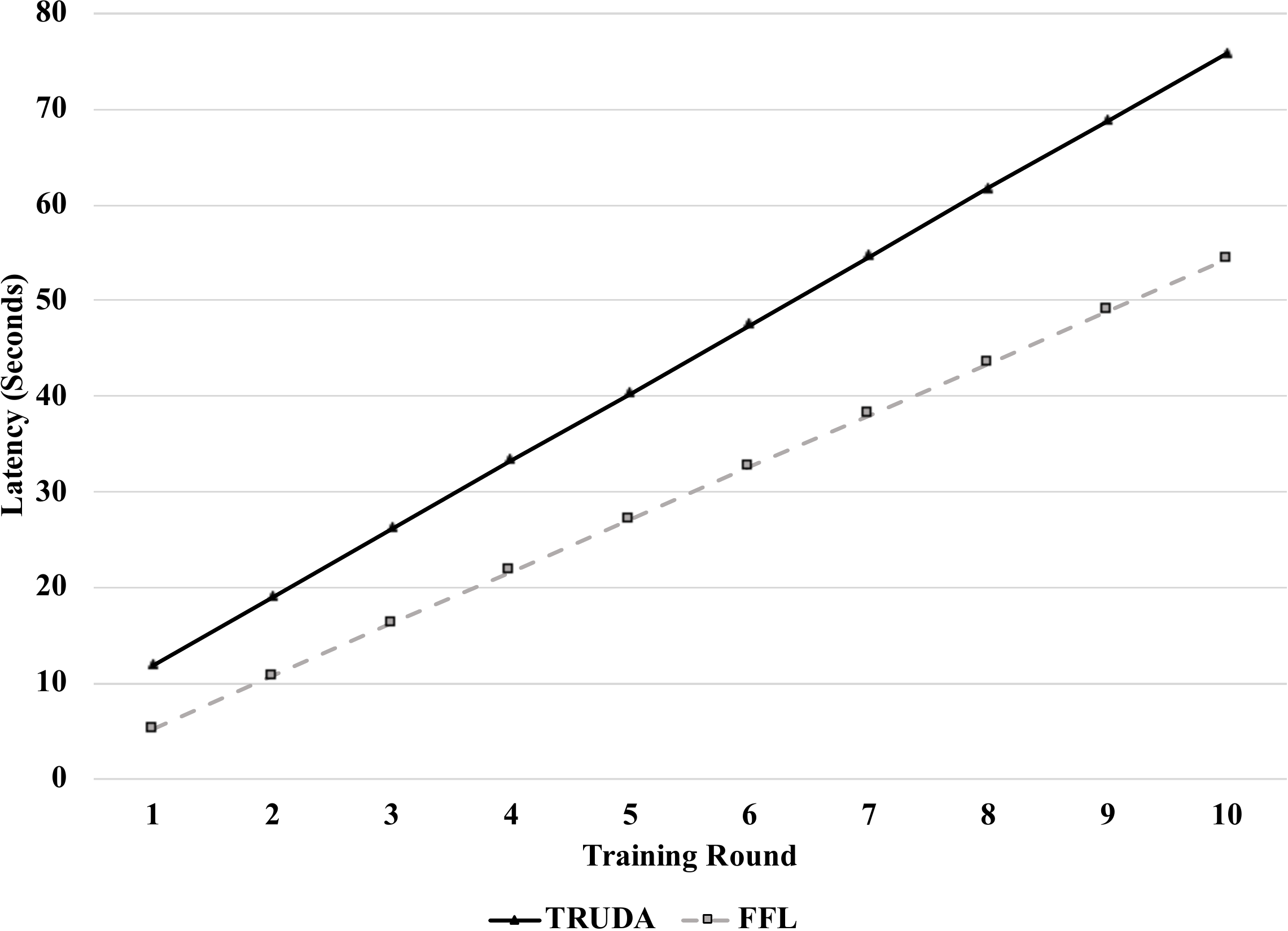}
\label{fig:mnistiterlatency}}
\subfloat[Latency Comparison: Coordinate Median]{
\includegraphics[width=0.33\textwidth]{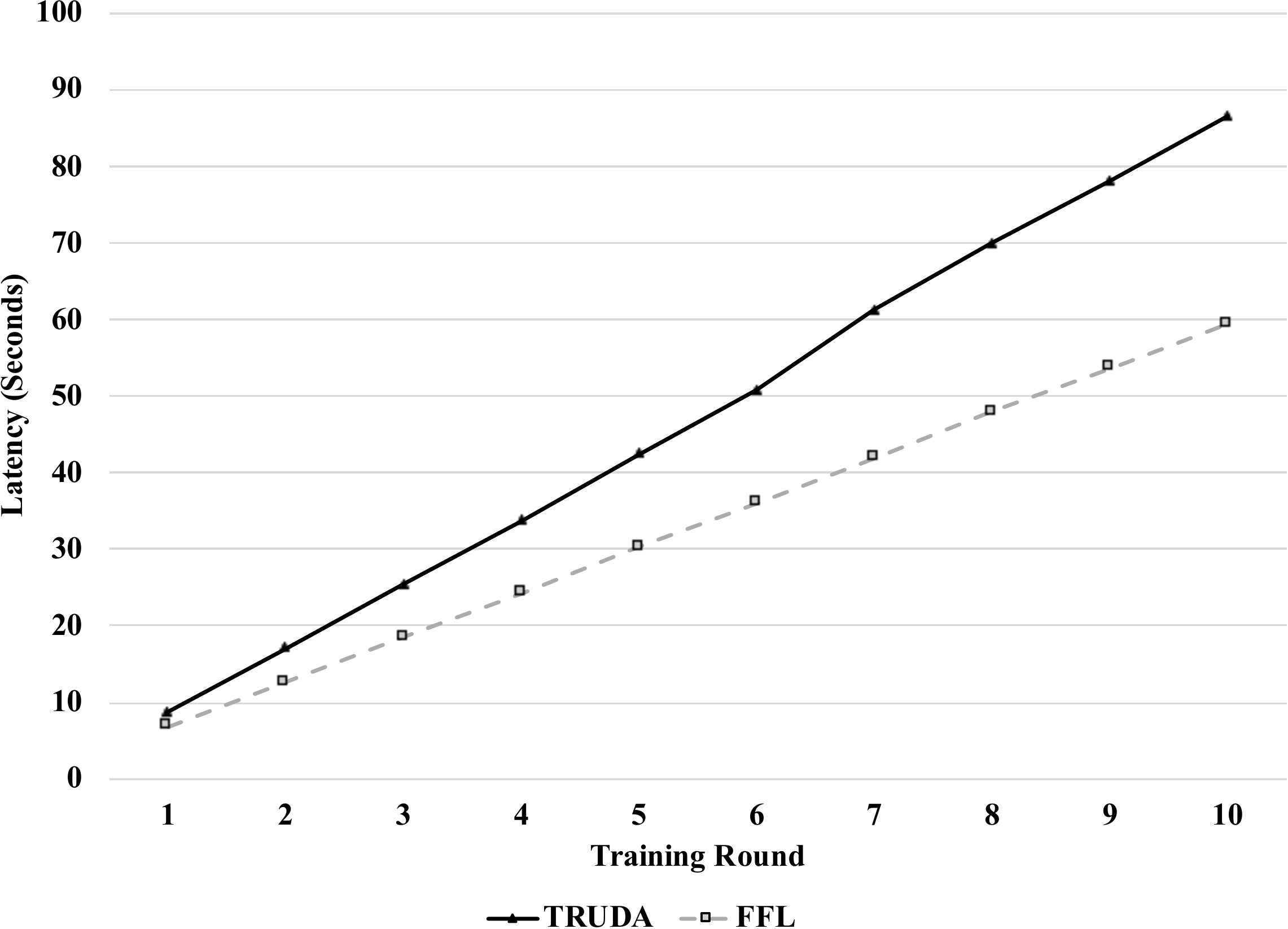}
\label{fig:mnistmedianlatency}}
\subfloat[Latency Comparison: Paillier crypto system]{
\includegraphics[width=0.33\textwidth]{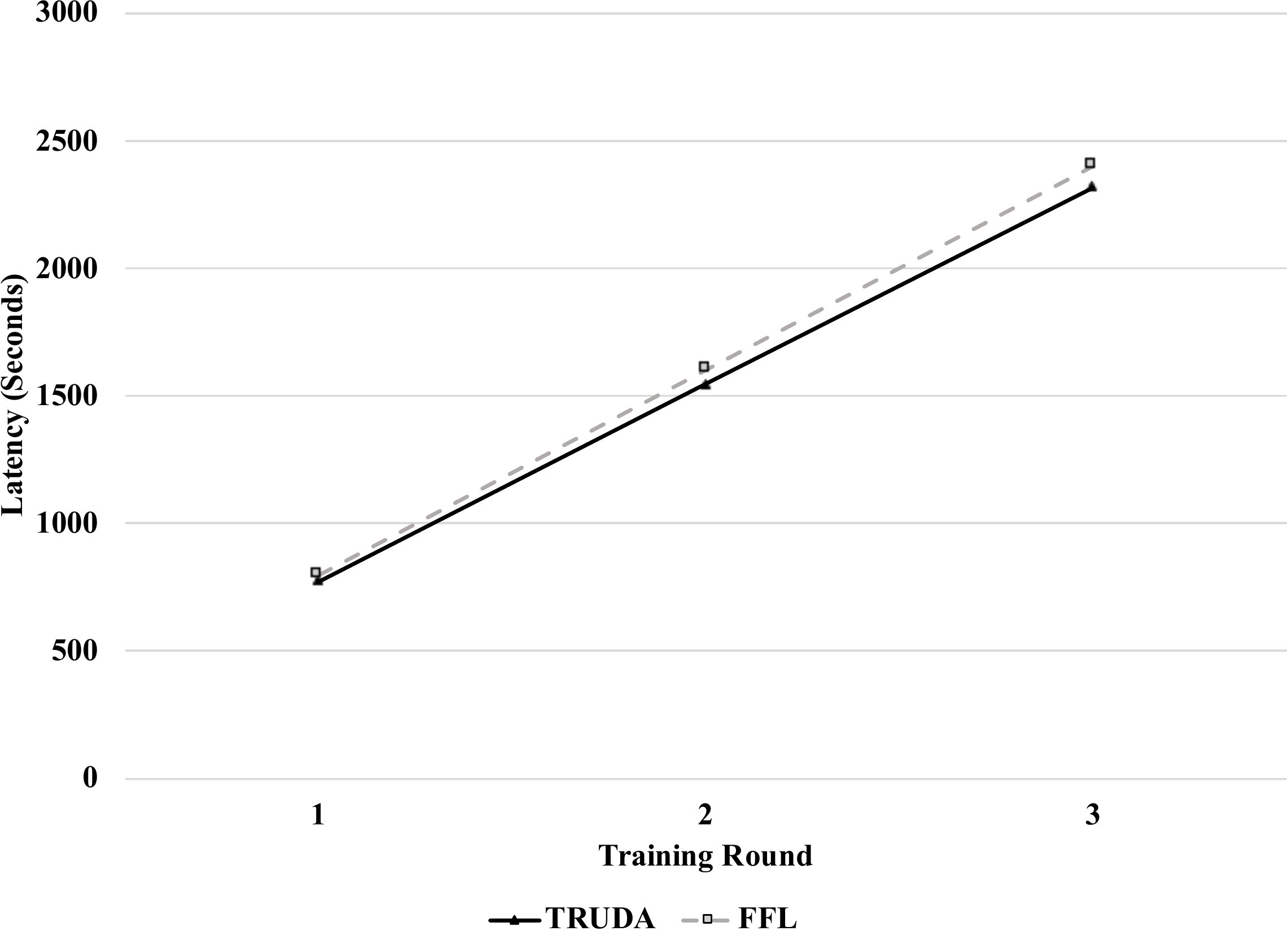}
\label{fig:mnistpaillierlatency}}
\caption{MNIST Loss/Accuracy/Latency Comparison Between \sysname{} and \ac{FFL} (IID with Four Parties)}
\label{fig:mnist}
\end{figure*}

As indicated in the \sysname{}'s design, we support aggregation algorithms with coordinate-wise arithmetic operations. We evaluated \sysname{} respectively with three aggregation algorithms, i.e., \emph{Iterative Averaging}, \emph{Coordinate Median}\cite{yin2018byzantine}, and \emph{Paillier}\cite{liu2019secure,truex2019hybrid}, with the \emph{MNIST} dataset. \emph{Iterative Averaging} is the base algorithm supporting \emph{\ac{FedAvg}} and \emph{\ac{FedSGD}}. It sends queries to all registered parties at each training round to collect information, e.g., model updates or gradients, averages the updates, and broadcasts the fused results to all parties. \emph{Coordinate Median} is a fusion algorithm that selects a coordinate-wise median from collected responses in order to tolerate Byzantine failures of adversarial parties. The \emph{Paillier crypto fusion} algorithm supports aggregation with Additively Homomorphic Encryption\cite{paillier1999public}. We trained deep learning models on the \emph{MNIST} dataset with ten training rounds for \emph{Iterative Averaging}, \emph{Coordinate Median} and three rounds for \emph{Paillier}. Each round has three local epochs.    

\emph{MNIST} contains $60,000$ examples in the training set. We randomly partitioned the training set into four equal sets for four parties. Each party has $15,000$ examples for local training. The trained model is a \ac{ConvNet} with eight layers. The detailed model architecture can be found in Table~\ref{tab:mnist} of Appendix~\ref{sec:trainingappendix}.

\nip{Accuracy/Loss and Convergence Rate.} 
We present the model loss and accuracy at each training round in Figures \ref{fig:mnistiterlossacc}, \ref{fig:mnistmedianlossacc}, and \ref{fig:mnistpaillierlossacc}. The horizontal axes are the number of training rounds. The left vertical axes show the loss and the right vertical axes present the model accuracy. It is clear that the loss/accuracy results of \sysname{} and \ac{FFL} have the same patterns for all three fusion algorithms. \sysname{} and \ac{FFL} converge at the same rate on \emph{MNIST} after one training round. The final models achieve the same accuracy level (above $98\%$) for both \sysname{} and \ac{FFL}.

\nip{Training Latency.}
We present the training latency data of \sysname{} and \ac{FFL} in Figures \ref{fig:mnistiterlatency}, \ref{fig:mnistmedianlatency}, and \ref{fig:mnistpaillierlatency}. The vertical axes are the accumulated time spent to finish that training round. We observed that for \emph{Iterative Averaging}, \sysname{} used $75.83$ seconds to finish the 10-round training and \ac{FFL} used $54.32$ seconds. Compared to the baseline \ac{FFL} system, the added security features in \sysname{} incurred additional $0.40\times$ latency for training the \emph{MNIST} model. Similarly for \emph{Coordinate Median}, \sysname{} incurred additional $0.45\times$ in latency. 

Due to heavyweight additively homomorphic encryption operations, \emph{Paillier crypto fusion} is two orders slower for training the same \emph{MNIST} model than \emph{Iterative Averaging} and \emph{Coordinate Median}. However, \sysname{} finished training with $0.04\times$ improvement in latency compared to \ac{FFL}. The reason is that the dominant performance factors of \emph{Paillier} fusion are the encryption/decryption operations. However, as the models are partitioned in \sysname{} for multiple decentralized aggregators, the \emph{Paillier} encryption/decryption and aggregation are accelerated --- computed in parallel by operating on smaller model partitions both on the aggregators and on the parties.    

\subsection{Training with Different Numbers of Parties}
\begin{figure*}[!ht]
\centering
\subfloat[Loss/Accuracy Comparison (4P vs. 8P)]{
\includegraphics[width=0.5\textwidth]{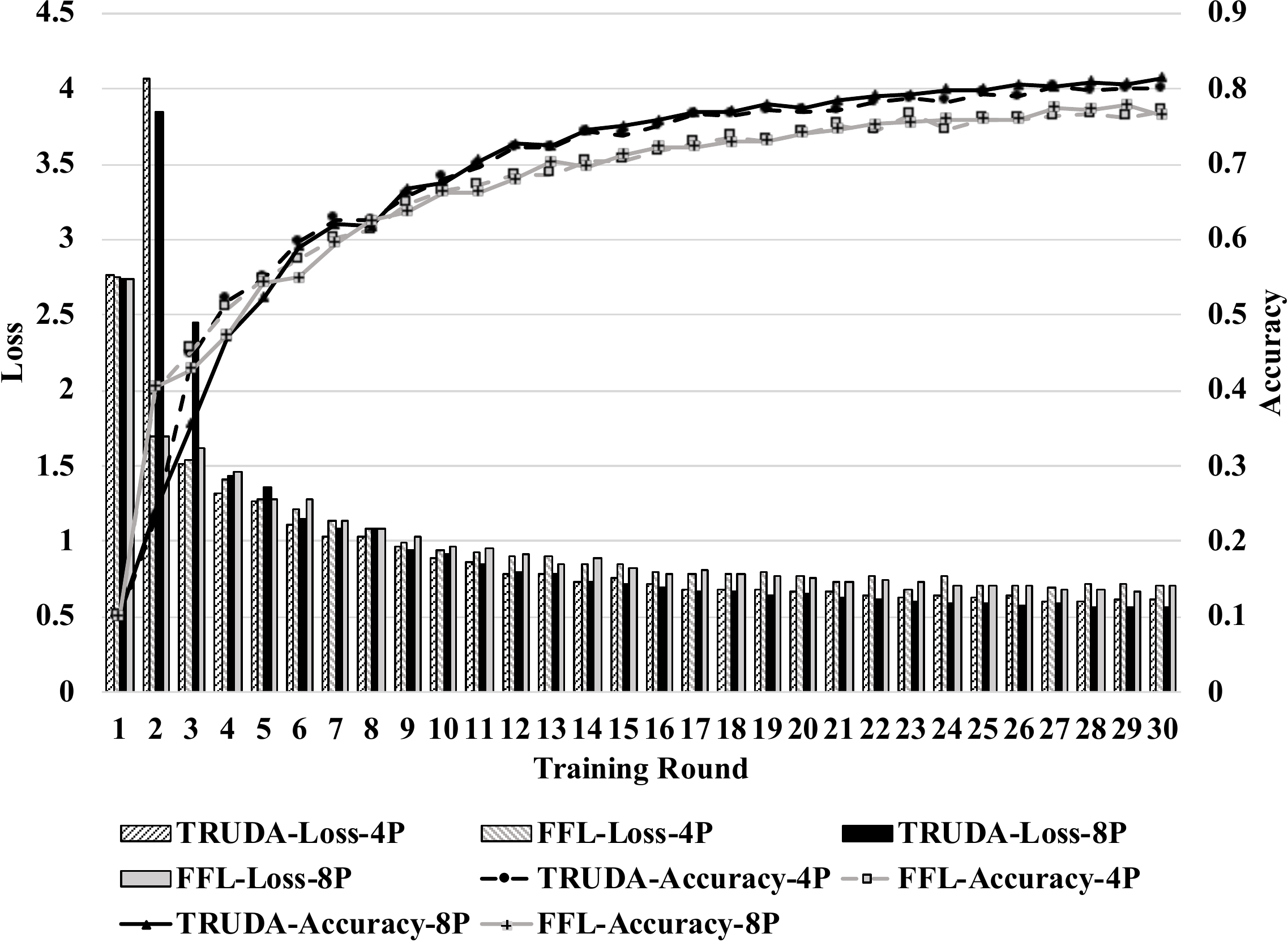}
\label{fig:cifar10lossacc}}
\subfloat[Latency Comparison (4P vs. 8P)]{
\includegraphics[width=0.5\textwidth]{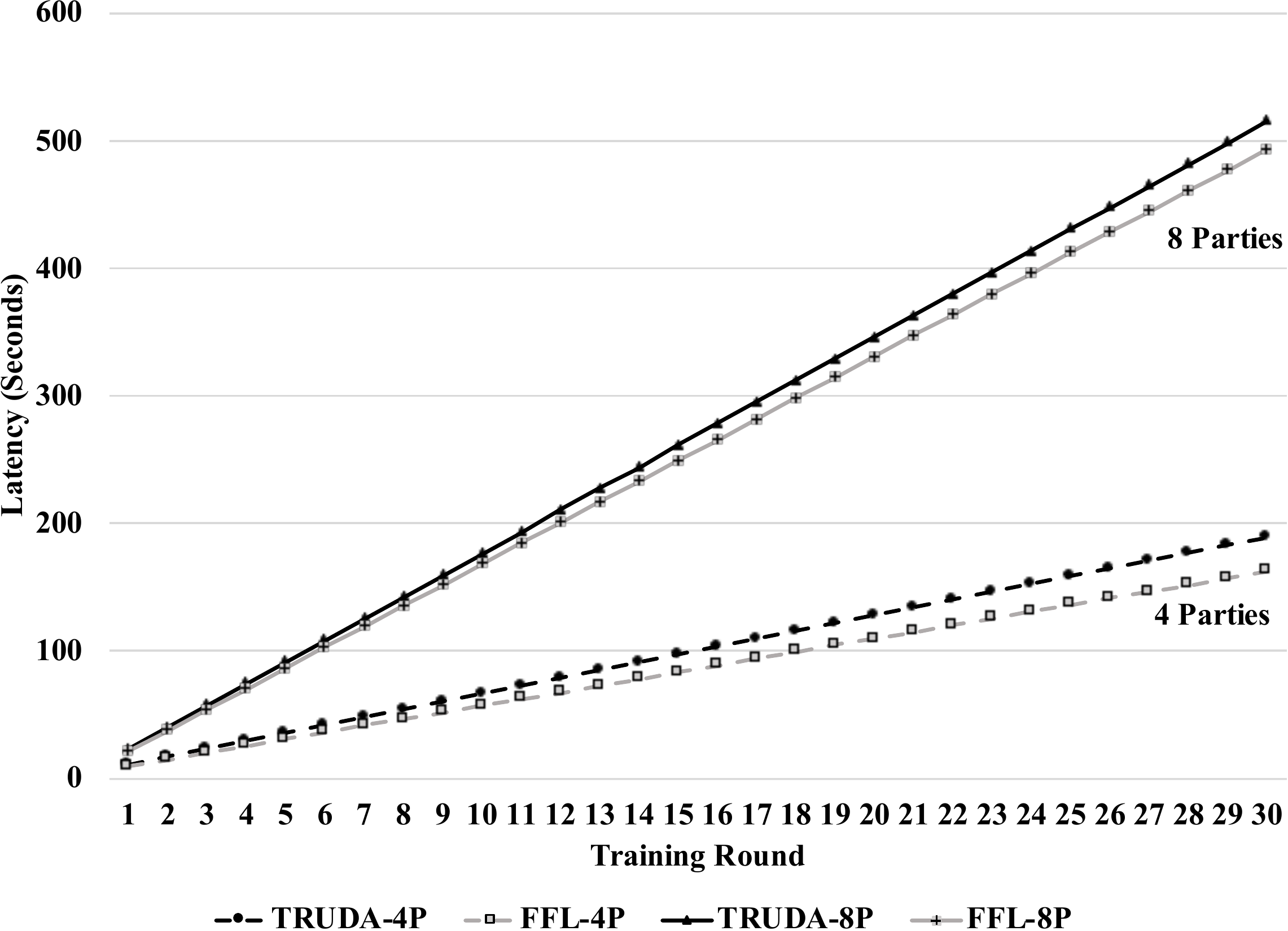}
\label{fig:cifar10latency}}
\caption{CIFAR10 Loss/Accuracy/Latency Comparison Between \sysname{} and \ac{FFL} (IID with Four/Eight Parties)}
\label{fig:cifar10}
\end{figure*}
\begin{figure*}[!ht]
\centering
\subfloat[Loss/Accuracy Comparison]{
\includegraphics[width=0.5\textwidth]{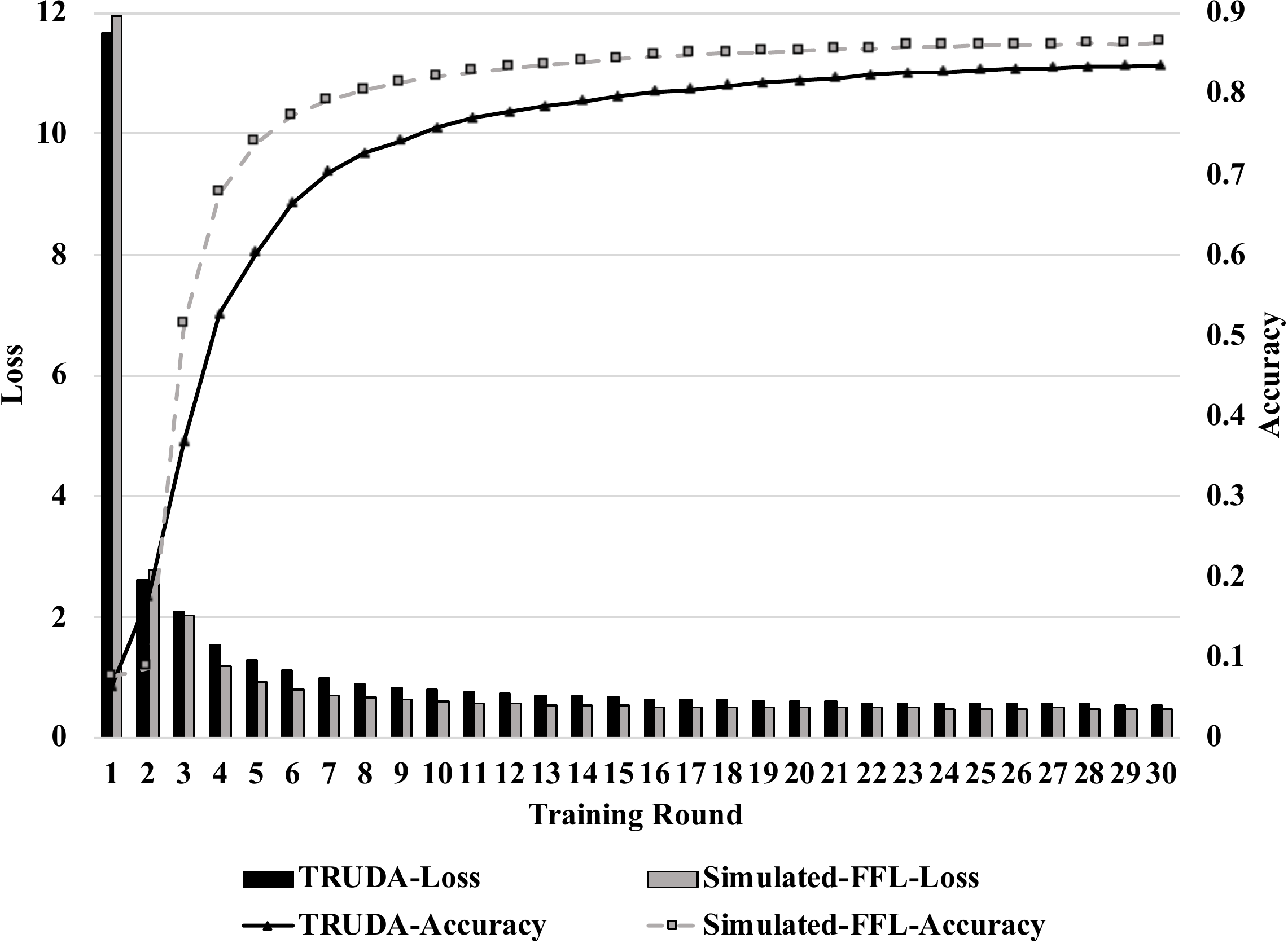}
\label{fig:rvlcdiplossacc}}
\subfloat[Latency Comparison]{
\includegraphics[width=0.5\textwidth]{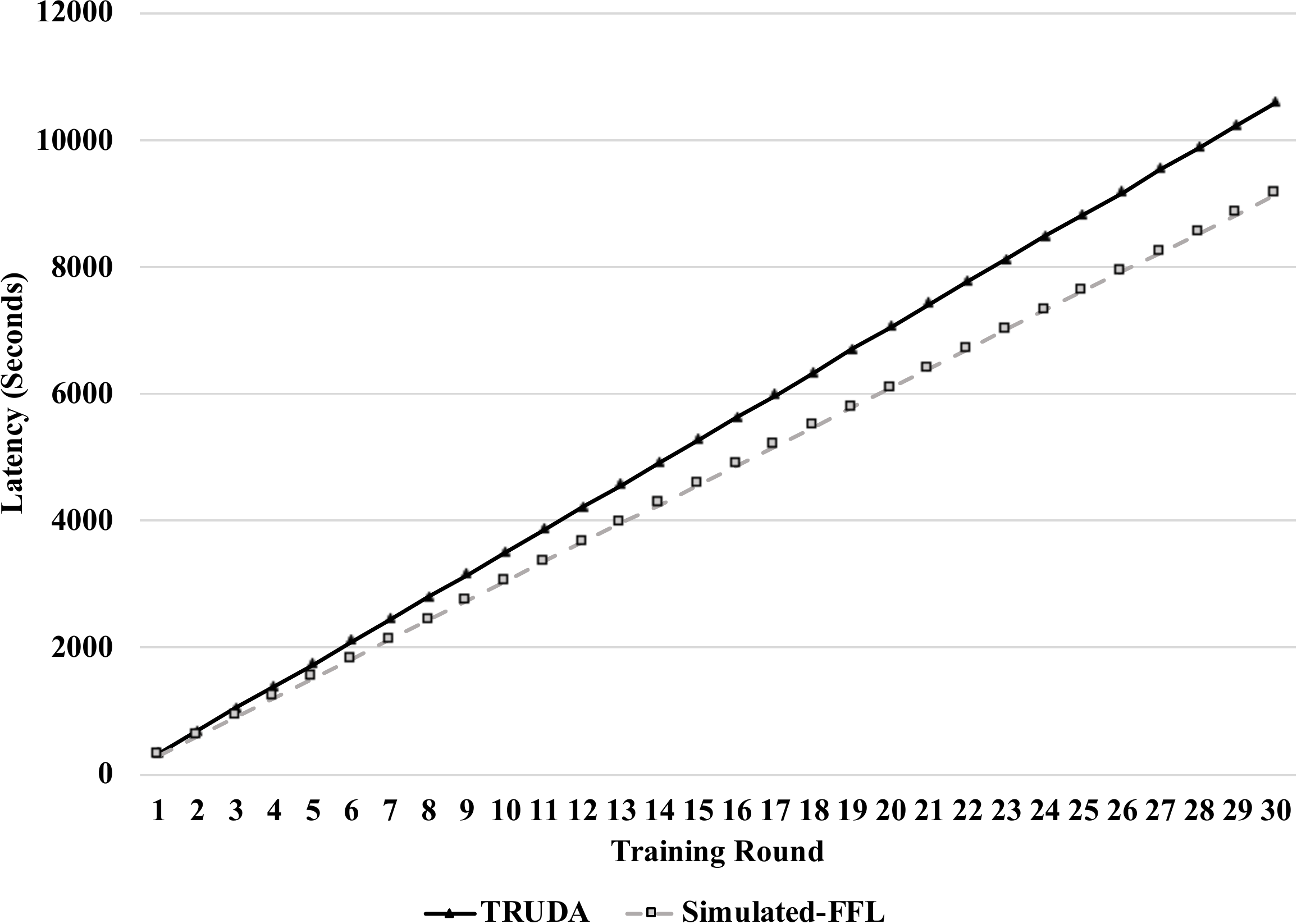}
\label{fig:rvlcdiplatency}}
\caption{RVL-CDIP Loss/Accuracy/Latency Comparison Between \sysname{} and \ac{FFL} (non-IID with Eight Parties)}
\label{fig:rvlcdip}
\end{figure*}
Here, we aim to understand the performance effects of involving more parties. 
We trained a \ac{ConvNet} with $23$ layers on \emph{CIFAR-10} with four and eight parties. 
The detailed model architecture can be found in Table~\ref{tab:cifar10} in Appendix~\ref{sec:trainingappendix}.
\emph{CIFAR-10} is a dataset with $32\times32$ color images in $10$ classes. We randomly partition the training set into equal sets for the parties. Each party has $10,000$ examples for \ac{FL} training. We trained this model with $30$ training rounds,
with each round consisting of one local epoch.

\nip{Accuracy/Loss and Convergence Rate.}
We present the model accuracy and loss at each training round in Figure~\ref{fig:cifar10lossacc}. The patterns for the loss and accuracy are similar for \sysname{} and \ac{FFL} with both four parties and eight parties. It indicates that the models converge at a similar rate with different number of parties. The accuracy results of the final model trained with \ac{FFL} are $76.99\%$ (four parties) and $76.43\%$ (eight parties). The final model accuracy results with \sysname{} are $79.93\%$ (four parties) and $81.41\%$ (eight parties).

\nip{Training Latency.}
We present the training latency data of \sysname{} and \ac{FFL} in Figure~\ref{fig:cifar10latency}. In the four parties scenario, it took \sysname{} $182.91$ seconds to finish $30$ rounds of training and \ac{FFL} $157.41$ seconds. Our added features incurred additional $0.16\times$ in latency for training the \emph{CIFAR-10} model. In the eight parties scenario, it took \sysname{} $498.68$ seconds to finish $30$ rounds of training and \ac{FFL} $477.34$ seconds. The latency only increased by $0.04\times$. We also find that adding more parties increases the latency for both \ac{FFL} and \sysname{} at the same pace. The security features of \sysname{} does not lead to additional latency with regard to more parties.

\subsection{Training with Non-IID Training Data}
We also measured the performance of training a larger, more complex deep learning model with non-IID training data distribution. We used a pre-trained \emph{VGG-16} model on the \emph{ImageNet} to train a document classifier on the \emph{RVL-CDIP}\cite{harley2015icdar} dataset with $16$ classes. For transfer learning with \emph{RVL-CDIP} classification, we replaced the last three fully-connected layers of \emph{VGG-16} due to differences in number of prediction classes. The detailed model architecture can be found in Table~\ref{tab:rvlcdip} in Appendix~\ref{sec:trainingappendix}.
The \emph{RVL-CDIP} dataset has $320,000$ training images and $40,000$ test images. We partitioned the training data based on the non-IID 90-10 skew data split for eight parties. Each party has approximately $40,000$ training examples with skew distribution among different classes, i.e., the two dominant classes contain $90\%$ of training data, while the remaining $14$ classes have $10\%$. We trained deep learning models with $30$ training rounds. Each round has one epoch. As \emph{RVL-CDIP} dataset it not officially supported in \ac{FFL}, we simulated the \ac{FFL} implementation for performance comparison.   

\nip{Accuracy/Loss and Convergence Rate.}
We present the model accuracy and loss at each training round in Figure~\ref{fig:rvlcdiplossacc}. Similarly, the models converge at a similar rate with \sysname{} and \ac{FFL}. The accuracy results of the final model trained with \sysname{} is $83.50\%$ and $86.19\%$ with simulated \ac{FFL}.

\nip{Training Latency.}
We present the training latency data of \sysname{} and simulated \ac{FFL} in Figure~\ref{fig:rvlcdiplatency}. It took \sysname{} $2.85$ hours and \ac{FFL} $2.46$ hours to finish $30$ rounds of training. Our added security features in \sysname{} incurred additional $0.16\times$ in latency for training the \emph{RVL-CDIP} model.  
\section{Related Work}
\label{sec:relate}
In this section, we give an overview of the security defenses against the \ac{FL} privacy leakage attacks and analyze their pros and cons from the perspectives of security and utility trade-off. In addition, we compare \sysname{} with them to demonstrate the  contributions of our work.

\subsection{\aclp{TEE}}
\acp{TEE} allow users to out-source their computation to third-party cloud servers with trust on the CPU package. 
They are particularly attractive for collaborative \ac{ML} computation, which may involve a large amount of privacy-sensitive training data, multiple distrusting parties, and stricter data protection regulations. 
\acp{TEE} can act as trustworthy intermediaries for isolating and orchestrating \ac{ML} processes and replace the expensive cryptographic primitives. 
For example, \ac{SGX} has been leveraged to support secure model inference~\cite{tramer2018slalom, gu2018confidential}, privacy-preserving multi-party machine learning~\cite{ohrimenko2016oblivious, hunt2018chiron, hynes2018efficient, gu2019reaching, mo2021ppfl}, and analytics on sensitive data~\cite{schuster2015vc3,bittau2017prochlo, zheng2017opaque,dave2020oblivious}. 

However, \acp{TEE} are not panacea to address all trust problems. Existing \ac{TEE} technologies have different performance and capacity constraints. For example, \ac{SGX} has a limit (128 MB) on the enclave's physical memory size. 1st-generation \ac{SEV} does not provide integrity protection. External \ac{ML} accelerators cannot be exploited for trusted execution with runtime memory encryption. In addition, \acp{TEE} may still be susceptible to emerging security vulnerabilities~\cite{van2020cacheout,van2020sgaxe,vanbulck2020lvi,murdock2020plundervolt,van2018foreshadow,werner2019severest,buhren2019insecure,li2019exploiting,li2020crossline,wilke2020sevurity}. One small defect may break the foundation of the entire trustworthy system. Thus, to assess a \ac{TEE}-integrated system, we need to distinguish security properties of different \acp{TEE} and extend the threat model in case of \ac{TEE} failure.
%Ohrimenko et al.~\cite{ohrimenko2016oblivious} employed Intel \ac{SGX} for privacy-preserving multi-party machine learning. They also proposed a series of data-oblivious \ac{ML} algorithms to prevent adversaries from learning the memory access patterns out of \ac{SGX} enclaves. 
%Chiron~\cite{hunt2018chiron} is an \ac{SGX}-based system for privacy-preserving \ac{MLaaS} and it also supports distributed training with multiple concurrent enclaves that exchange model parameters via a parameter server. Myelin~\cite{hynes2018efficient} is another research work that leverages \ac{SGX} for private machine learning on multi-source private data. 
%Slalom~\cite{tramer2018slalom} explored out-sourcing \ac{DNN}'s linear computation to faster GPUs with integrity verification. Gu et al.~\cite{gu2018confidential} investigated partitioning of deep learning models and offloading insensitive workloads for hardware acceleration. 
%CalTrain~\cite{gu2019reaching} leveraged \ac{TEE} to address the problem that preserving data confidentiality is in conflict with model accountability in collaborative learning. 
%Graviton~\cite{volos2018graviton} intended to extend GPU with crypto engines and build secure channels between the \ac{SGX} enclaves and GPUs. Thus, GPUs can be incorporated into the trust domain. 
%HIX~\cite{jang2019heterogeneous} built a GPU enclave with exclusive access to GPUs and all other enclaves had to route their workloads through the GPU enclave. Thus, they can enforce access control to GPUs for trusted execution. 

Sharing a similar goal with all the research works mentioned above, we leverage \acp{TEE} to support trustworthy aggregation as the first-line defense against privacy leakages. The unique contributions of our work are to decentralize the aggregation to multiple trusted execution entities and employ dynamic permutation to further obfuscate the model updates. Thus, we are still resilient to data reconstruction attacks even if model updates are leaked from breached \acp{TEE}. 

\subsection{Cryptographic Schemes}
\acl{HE} allows arithmetic operations on ciphertexts without decryption.   
Aono et al.~\cite{aono2017privacy} used additively \ac{HE} to protect the privacy of gradients to prevent information leakage. 
Hardy et al.~\cite{hardy2017private} encrypted vertically partitioned data with an additively \ac{HE} and learned a linear classifier in the \ac{FL} setting. 
\acl{SMC} allows different parties to compute a joint function without revealing their inputs to each other and has been widely researched in collaborative analytics and multi-party learning~\cite{mohassel2017secureml,bonawitz2017practical,zheng2019helen, zheng2021cerebro, poddar2021senate,truex2019hybrid, bonawitz2017practical}.

%SecureML~\cite{mohassel2017secureml} conductedprivacy-preserving learning via \ac{SMC}. It required data owners to encrypt and secretly share their data among two non-colluding servers in the initial setup.
%Helen~\cite{zheng2019helen} is a coopetitive system using maliciously \ac{SMC} for training linear models with a strong adversarial setting where each party could only trust itself.
%Bonawitz et al.~\cite{bonawitz2017practical} proposed a secure aggregation protocol for aggregating individual model updates via \ac{SMC}. The servers can only learn the information from the aggregated results. It is also robust when clients frequently drop in the \ac{FL}'s cross-device setting. 

Compared to \ac{TEE}-based approaches like \sysname{}, using cryptographic schemes for privacy protection promise same-level privacy/accuracy and can exclude CPU vendors from the trust domain. But the trade-offs are the extra communication and computational overhead. The performance constraints for now may still hinder their applications for large-scale \ac{FL} training scenarios. 

\subsection{\acl{DP}}
In the \ac{ML} setting, \ac{DP} can be used to apply perturbations for mitigating information leakage. Compared to cryptographic schemes, \ac{DP} is more computationally efficient at the cost of a certain utility loss due to the added noise. 
\ac{CDP} data analysis is typically conducted under the assumption of a trusted central server. \ac{CDP}~\cite{mcmahan2017learning, geyer2017differentially} can achieve an acceptable balance between privacy and accuracy, but it does not fit a threat model where the aggregation server might be honest-but-curious, malicious, or compromised.
In an effort to remove the trusted central server assumption in the threat model, \ac{LDP}~\cite{shokri2015privacy, bhowmick2018protection} lets each client conduct differentially private transformations of their private data before sharing them with the aggregator. However, achieving \ac{LDP} comes at the cost of utility loss as every participant must add enough noise to ensure \ac{DP} in isolation. 

Due to conflicting threat models, the design of decentralized and shuffled aggregation in \sysname{} is incompatible with \ac{CDP}, which requires a central aggregator in \ac{FL} training. However, we can still enable trustworthy aggregation to strengthen \ac{CDP} with a \ac{TEE}-protected aggregator. \sysname{} can be seamlessly integrated with \ac{LDP} as the \ac{LDP}'s perturbations only apply to model updates on the party machines.    
\section{Conclusion}
\label{sec:conclusion}
In light of the recent training data reconstruction attacks targeting \ac{FL} aggregation, we rethink the trust model and system architecture of \ac{FL} frameworks and identify the root causes of their susceptibility to such attacks. To address the problem, we exploit the unique computational properties of aggregation algorithms and propose protocol/architectural enhancements to minimize the leakage surface and break the information concentration. We have developed \sysname{}, a new cross-silo \ac{FL} system encompassing three security-reinforced mechanisms, i.e., trustworthy, decentralized, and shuffled aggregation. Therefore, each aggregator only has a fragmentary and obfuscated view of the model updates. We demonstrate that \sysname{} can effectively mitigate all training data reconstruction attacks with no utility loss and low performance overheads. 
\balance
\bibliographystyle{ACM-Reference-Format}
\bibliography{main}
\newpage
\appendix
\begin{figure*}[h]
\centering
\includegraphics[width=1\textwidth]{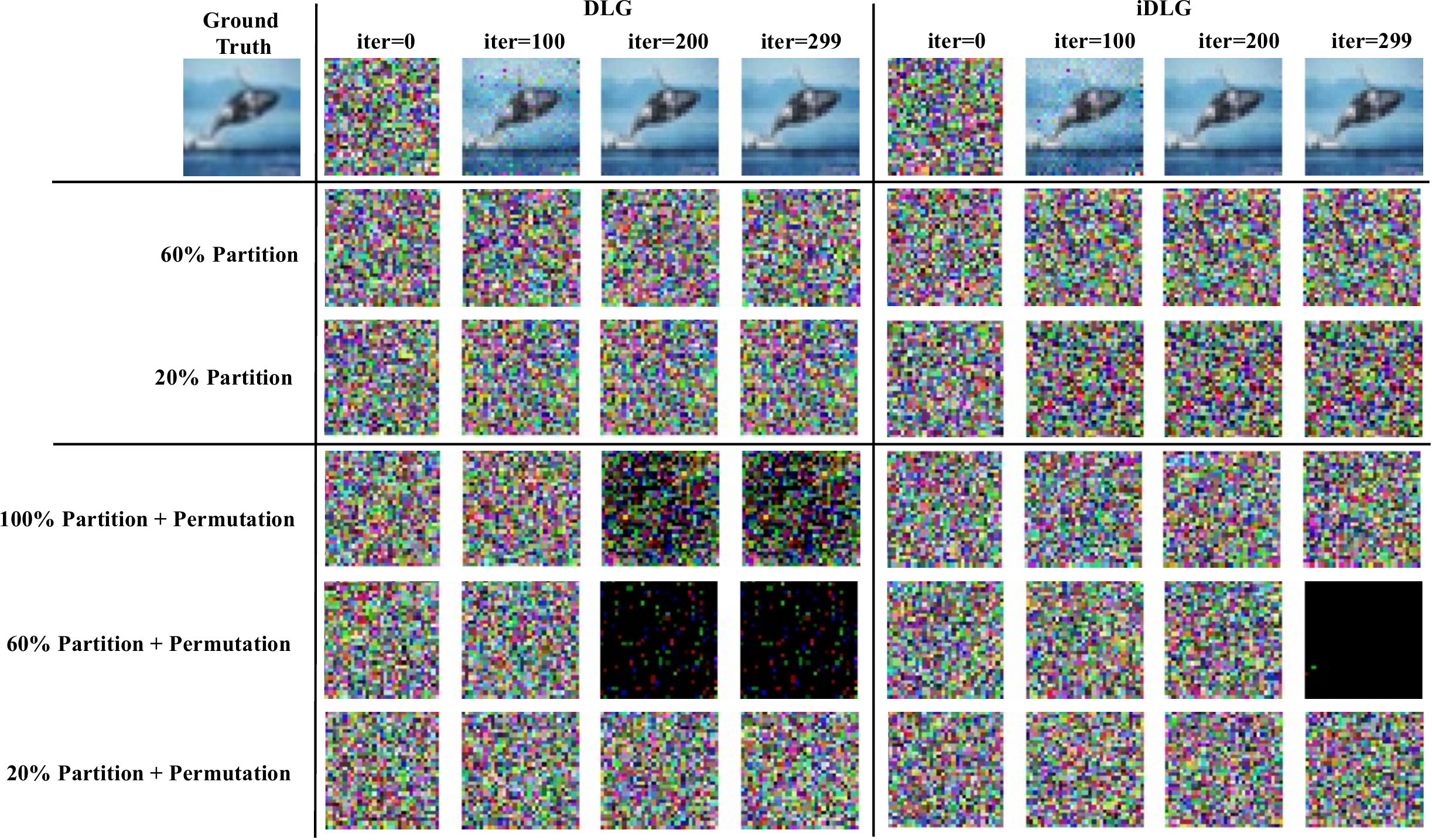}
\caption{Reconstruction Examples of DLG and iDLG with Model Partitioning and Permutation}
\label{fig:dlgidlgexamples}
\end{figure*}

\section{Tables for DNN Architectures in FL Training}
\label{sec:trainingappendix}
Here we present the detailed model architectures and hyper-parameters for the three \acp{DNN} we trained in Section~\ref{sec:performance}.
The \emph{Layer} column shows the layer types,
including convolutional layer (conv), dense layer (dense), flatten layer (flat), max pooling layer (maxpool), batch normalization layer (batchnorm), and dropout layer (dropout). 
The \emph{Filter} column shows the number of filters in each
convolutional layer. The \emph{Size} column is in the format of 
\emph{width} $\times$ \emph{height} to represent filter parameters. The \emph{Activation} column shows the type of activation function used. 

\begin{table}[!ht]
\centering
\caption{\ac{DNN} Architecture for \emph{MNIST}}
\label{tab:mnist}
\begin{tabular}{lccc}
\hline
Layer     & Filter & Size & Activation             \\ \hline
1 conv    & 32     & 3x3   & relu\\
2 conv    & 64     & 3x3   & relu\\
3 maxpool &        & 2x2         \\
4 dropout  & \multicolumn{2}{c}{p = 0.25}                      \\
5 flatten                        \\
6 dense   & 128    &       & relu\\ 
7 dropout     & \multicolumn{2}{c}{p = 0.50}                   \\
8 dense   & 10     &       & softmax\\\hline
\end{tabular}
\end{table}

\begin{table}[!b]
\centering
\caption{\ac{DNN} Architecture for \emph{CIFAR-10}}
\label{tab:cifar10}
\begin{tabular}{lccc}
\hline
Layer     & Filter & Size & Activation             \\ \hline
1 conv    & 32     & 3x3   & relu\\
2 batchnorm                      \\
3 conv    & 32     & 3x3   & relu\\
4 batchnorm                      \\
5 maxpool &        & 2x2         \\
6 dropout & \multicolumn{2}{c}{p = 0.20} \\
7 conv    & 64     & 3x3   & relu\\
8 batchnorm                      \\
9 conv    & 64     & 3x3   & relu\\
10 batchnorm                      \\
11 maxpool &        & 2x2         \\
12 dropout & \multicolumn{2}{c}{p = 0.30} \\
13 conv    & 128     & 3x3   & relu\\
14 batchnorm                       \\
15 conv    & 128     & 3x3   & relu\\
16 batchnorm                       \\
17 maxpool &        & 2x2         \\
18 dropout & \multicolumn{2}{c}{p = 0.40} \\
19 flatten                        \\
20 dense   & 128    &       & relu\\ 
21 batchnorm                       \\
22 dropout     & \multicolumn{2}{c}{p = 0.50}\\
23 dense   & 10     &       & softmax\\\hline
\end{tabular}
\end{table}

\begin{table}[b]
\centering
\caption{\ac{DNN} Architecture for \emph{RVL-CDIP}}
\label{tab:rvlcdip}
\begin{tabular}{lccc}
\hline
Layer     & Filter & Size & Activation             \\ \hline
1 conv    & 64     & 3x3   & relu\\
2 conv    & 64     & 3x3   & relu\\
3 maxpool &        & 2x2         \\
4 conv    & 128     & 3x3   & relu\\
5 conv    & 128     & 3x3   & relu\\
6 maxpool &        & 2x2         \\
7 conv    & 256     & 3x3   & relu\\
8 conv    & 256     & 3x3   & relu\\
9 conv    & 256     & 3x3   & relu\\
10 maxpool &        & 2x2         \\
11 conv    & 512     & 3x3   & relu\\
12 conv    & 512     & 3x3   & relu\\
13 conv    & 512     & 3x3   & relu\\
14 maxpool &        & 2x2         \\
15 conv    & 512     & 3x3   & relu\\
16 conv    & 512     & 3x3   & relu\\
17 conv    & 512     & 3x3   & relu\\
18 maxpool &        & 2x2         \\
19 flatten                        \\
20 dense   & 128    &       & relu\\ 
21 dropout     & \multicolumn{2}{c}{p = 0.50}\\
22 dense   & 16     &       & softmax\\\hline
\end{tabular}
\end{table}

\newpage
\section{Reconstruction Examples of DLG/iDLG/IG}
\label{sec:exampleappendix}

\begin{figure*}[h]
\centering
\includegraphics[width=0.9\textwidth]{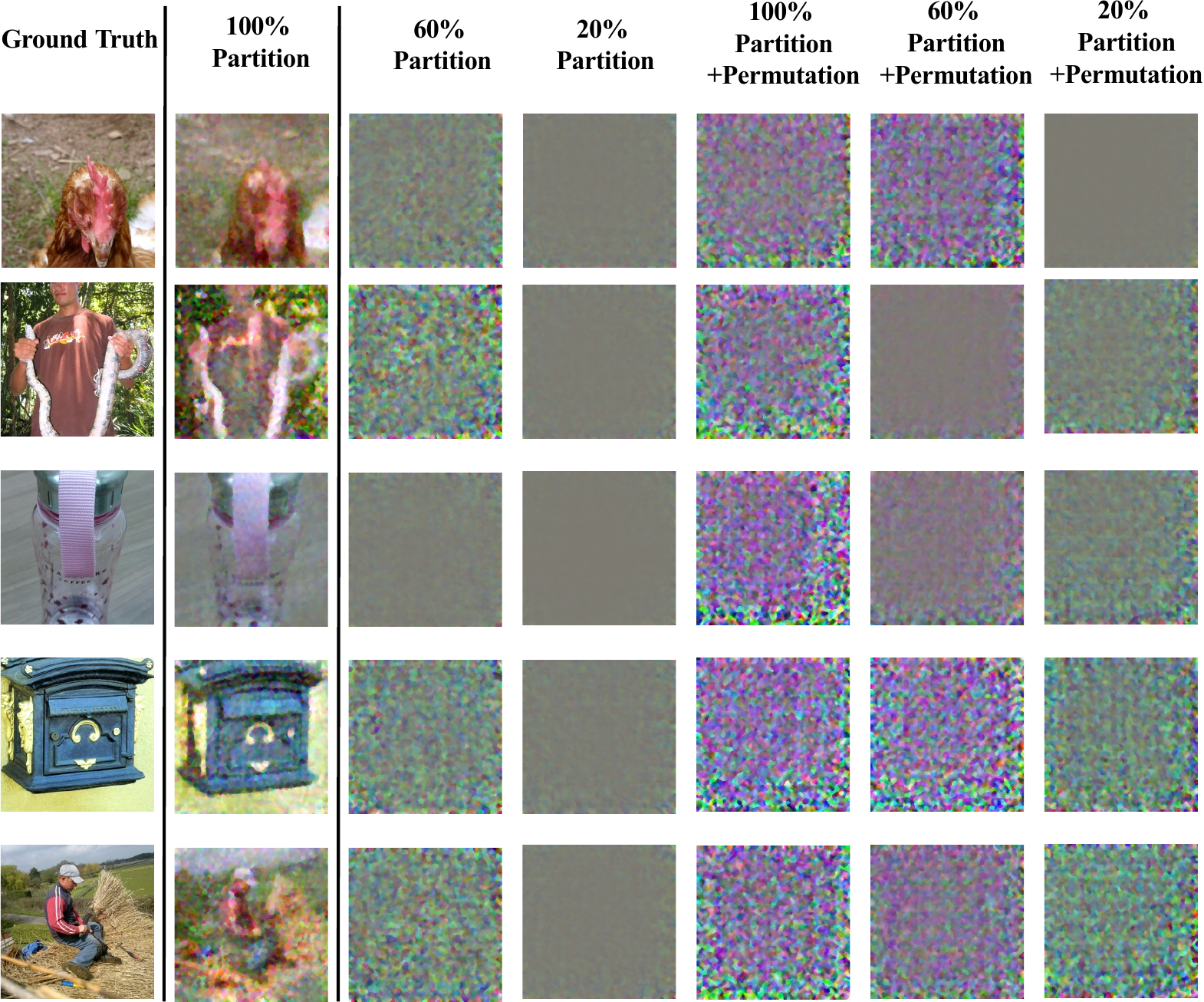}
\caption{Reconstruction Examples of \ac{IG} with Model Partitioning and Permutation}
\label{fig:igexamples}
\end{figure*}

In Figure~\ref{fig:dlgidlgexamples}, we display the intermediate results of reconstructing an image using \ac{DLG} and \ac{iDLG} with different combinations of partitioning and permutation parameters. The first row of images shows the reconstruction results with the entire model updates, i.e., $100\%$ partition, over $300$ iterations. We use it as the baseline for comparison. The images at iteration $0$ is the randomly initialized input. We observed that with the entire, unperturbed model update, \ac{DLG} and \ac{iDLG} can reconstruct most of the original image after $100$ iterations.    
In the second and third row, we randomly kept $60\%/20\%$ of the model updates and provided them to the \ac{DLG} and \ac{iDLG} attacks. With only partial information, both attacks were unable to correctly minimize their respective cost function. Thus, the intermediate and final reconstructions do not contain any recognizable information related to the ground truth example. 
In the fourth through sixth rows, we randomly permuted the model updates with $100\%/60\%/20\%$ partitions respectively and ran the \ac{DLG} and \ac{iDLG} attacks. Similar to the previous results with partitioning only, the intermediate reconstructions do not contain any recognizable information related to the ground truth example. Of note is the fourth row of images, as they demonstrate that \sysname{}'s permutation mechanism works even when only a single aggregator is present, which is traditionally used in \ac{FL}. 

In Figure~\ref{fig:igexamples}, we display five \emph{ImageNet} examples used in our \ac{IG} reconstruction experiments. The first column shows the original images. The second column presents the reconstructed images (after $24,000$ iterations) when the entire model updates ($100\%$ partition) were provided to the \ac{IG} attack. These results are used as the baseline for comparison. In the third through seventh columns, we present the reconstructions with different combinations of partitioning and permutation enabled. Compared to the baseline, none of the reconstructions contain recognizable information related to the ground truth examples. Without the complete, unperturbed model updates, \ac{IG} is unable to minimize its cost function.
\end{document}